# STC: Semantic Taxonomical Clustering for Service Category Learning

Sourish Dasgupta, Satish Bhat, and Yugyung Lee

***Abstract*— Service discovery is one of the key problems that has been widely researched in the area of Service Oriented Architecture (SOA) based systems. Service category learning is a technique for efficiently facilitating service discovery. Most approaches for service category learning are based on suitable similarity distance measures using thresholds. Threshold selection is essentially difficult and often leads to unsatisfactory accuracy. In this paper, we have proposed a self-organizing based clustering algorithm called *Semantic Taxonomical Clustering* (STC) for taxonomically organizing services with self-organizing information and knowledge. We have tested the STC algorithm on both randomly generated data and the standard OWL-S TC dataset. We have observed promising results both in terms of classification accuracy and runtime performance compared to existing approaches.**

## I. Introduction

ONE of the major operations in Service Oriented Architecture (SOA) based systems is service discovery. In order to facilitate dynamic on-demand access to services, we need an efficient way of discovering the required services out of a large pool of functionally different services. The service discovery process can become very efficient when service registries are categorically organized into well-defined access structures such as Universal Description Discovery & Integration (UDDI) [1] and Distributed Hash Tables (DHT) based registries such as CHORD [2]. A conventional way of grouping services into categories is to apply machine learning techniques for learning service categories.

The general problem statement is as follows: Given a set of service descriptions that contain functional and QoS features, we need to model a learner (system) that can predict the labeled or unlabeled category (mostly functional) of the services by observing their corresponding descriptions with a minimum prediction error.

A functionally similar category (i.e. clusters) can then be indexed into centralized registries or into DHTs depending upon the application, domain, and underlying network issues. A consumer query is then mapped over a cluster space and the service having the best match is selected. The evaluation of the best match is usually based upon a pre-defined ranking function and this process is called *Service Selection*. However, service category learning based on such statistical techniques suffers from some major disadvantages (as will be discussed in Section 3). One of these is the problem of optimal threshold selection that is innate for all pair-wise comparison of service descriptions during the learning process. Another important drawback is the inaccuracy caused due to lack of utilization of the semantic definitions of the terms that constitute the service descriptions.

In this paper, we propose a novel service category learning algorithm called *Semantic Taxonomical Clustering* (STC) that utilizes semantic descriptions of services. However, the STC approach does not apply statistical learning techniques (as is the case in most service category learning algorithms). The assumption behind this approach is that service descriptions have to be semantically defined using a set of domain ontologies. Within this framework service descriptions can be organized into a set of *taxonomies* where services with generic functionality are clustered at lower depths of the taxonomies while more specific services are clustered at higher depths. STC does not involve any distance measure over Euclidean space and hence, does not require selection of a threshold for clustering. Rather the clustering is self-organizing and the topology of the cluster space remains the same for a fixed set of service descriptions irrespective of the order in which the pair-wise comparisons are made.

The paper is organized as follows. In Section 2, we start with a discussion on some significant statistical service category learning approaches. Since all these approaches are based on the notion of a distance measure for similarity between services, Section 3 shows some of the limitations of distance measure based approaches in general. In Section 4, the STC algorithm together with its conceptual foundations and properties is proposed and analyzed. Section 5 presents a comparative empirical result that shows: (a) the runtime efficiency of the proposed algorithm as compared to a nearest neighbor based clustering algorithm over a set of randomly

Manuscript received December 3, 2011. This work was supported in part by the U.S. National Science Foundation under Grant IIS #0742666.

Sourish Dasgupta was with the University of Missouri, Kansas City, MO 64110 USA. He is now with the Department of Computer Science, DA-IICT University, Gandhi Nagar, Gujarat, India (e-mail: sourish_dasgupta@daiict.ac.in).

Satish Bhat was with the University of Missouri, Kansas City, MO 64110 USA. He is now with Machine Learning at Adknowledge, Kansas City, MO 64112 USA (e-mail: satish.ssb@gmail.com).

Yugyung Lee is with the University of Missouri, Kansas City, MO 64110 USA (corresponding author to provide phone: 816-235-5932; fax: 816-234-5932; e-mail: leeyu@ umkc.edu).

generated dataset and (b) the accuracy of STC when compared against an expert evaluated categorization of the standard OWLS-TC dataset. Section 6 discusses the limitation of the STC algorithm and suggestions for future research and Section 7 concludes this paper.

## II. Related Work

There have been several research works on service category learning so far. Service categorization is usually motivated by the thematic properties that have been proposed in standards such as UNSPCS (United Nations Standard Products and Service Codes) [3], NAICS (North American Industry Classification System) [4]. Thematic properties may include the service functional properties (input, output, pre-condition, and result), the QoS properties (availability, reliability, etc.), and domain information (i.e. area of application) that can be extracted out of service descriptions. A distance measure (either keyword-based [5 - 8] or ontology-based [9 - 13]) is modeled and used to compute the pair-wise similarity between two service descriptions. In key word based distance measures, the similarity of two services is computed based on the TF/IDF technique derived from IR research [5 - 8]. TF/IDF is done to ascribe weight to the documents (service descriptions) with respect to a particular term (attribute keyword). The weighted attributes (functional attributes are input, output, pre-condition, post-condition) of the services are represented as a feature vector and then the similarity between the attributes are computed based on conventional vector cosine similarity measure. IR based similarity computation can be very useful where we do not have formalized semantics for the service descriptions. As an alternative approach, in an ontology based semantic distance measure [9 - 13], ontological concepts having semantic definitions are used instead of syntactic tokens. The semantic distance measure over these concepts can be classified into three categories: (i) taxonomic distance based [14 - 15], (ii) information content (IC) based [16 – 17], and (iii) concept property based [18 – 20].

In most research works (as has been referred below) relating to service categorization the learning problem is limited to functional properties only. Service category learning is primarily done for service discovery that is primarily matching consumer requested service functionality with available services. In general, we can classify all such learning techniques into two basic learning frameworks: (i) supervised learning and (ii) unsupervised learning. In the supervised learning mode, it is essential to have a sufficiently "sound" training data of service descriptions that guarantees minimum over-fitting and under-fitting. Also, we need to have a clear understanding of the categories into which new service descriptions can be fitted. As mentioned before, research works have involved classical Machine Learning (ML) techniques such as SVMs (Support Vector Machine) [21 – 22] and NBC (Naïve Bayesian Classifier) [23 - 24]. Some works have also used more recent Information Retrieval models such as LSI (Latent Semantic Index) [25] and PLSI (Probabilistic LSI) [26 - 27].

The unsupervised learning mode is another alternative method of service category learning. In this mode we do not need to have pre-understanding of the service categories into which services have to be fitted. The training dataset is used to generate a learning model that is basically a set of clusters of service descriptions such that it maximizes the global inter-cluster distance (i.e., service functional dissimilarity) while minimizes the global intra-cluster distance. In other words, services are grouped into functionally similar clusters in such a way that new observations do not disturb the cluster space topology by re-modeling the learner (creating new clusters or modifying old clusters). One of the most common techniques in this direction of service category learning is using K-means based algorithms [28 - 29]. These algorithms are partitional in nature in the sense that they partition the training set into disjoint partitions (i.e. clusters) where the number of partitions is pre-estimated. Subsequent new service descriptions are then fitted into these partitions with minimum errors. Another technique of unsupervised learning is to use hierarchical clustering algorithms such as agglomerative based clustering [5, 9, 29]. In these approaches service descriptions are pair-wise compared to form a type of minimum spanning tree over the cluster space (instead of disjoint partitions). The tree structure enforces a partial ordering over the cluster space by representing nodes at lower depths to be a more generic set of similar services while nodes at higher depths are more specific sets of similar services. The partial order is essentially the intra-set distance that is lower in bottom level nodes while higher in top level nodes. In this approach, there is no requirement to pre-estimate the total number of nodes (i.e. sets/clusters) as they are self-induced by the algorithms.

## III. Shortcomings of Distance-based Learning

In general, most service category learning techniques (supervised and unsupervised) discussed so far are distance measure (i.e. similarity measure) based. In this section we discuss some of the major limitations of distance-based learning:

***Problem of Threshold Selection*** [5, 24]: Threshold selection is necessary for learning algorithms that require some sort of selection condition for two services to be considered similar. Barring a few techniques (such as k-means and k-nearest neighbor based algorithms), most service category learning algorithms require an optimal threshold selection. If the threshold is too tight then it might affect the recall while if the threshold is too loose then it might affect the precision. In most cases the choice of threshold is empirically done. This consumes the overall learning time period and requires a lot of manual intervention.

***Problem of Sample Selection Order for Online Learning*****:** In an online learning mode, we do not have a fixed service set to begin with. In such a framework the order in which services are observed and categorized, may have negative side effect on the overall clustering performance (e.g., KNN [24]). We call this a problem of sample selection order. To explain this problem, we take an example domain (Fig. 1). Let us consider three services $s_1$, $s_2$, and $s_3$ that need to be clustered according to their output feature O. It is given that $s_1.o$ = {*car, location*}, $s_2.o$ = {*vehicle,*

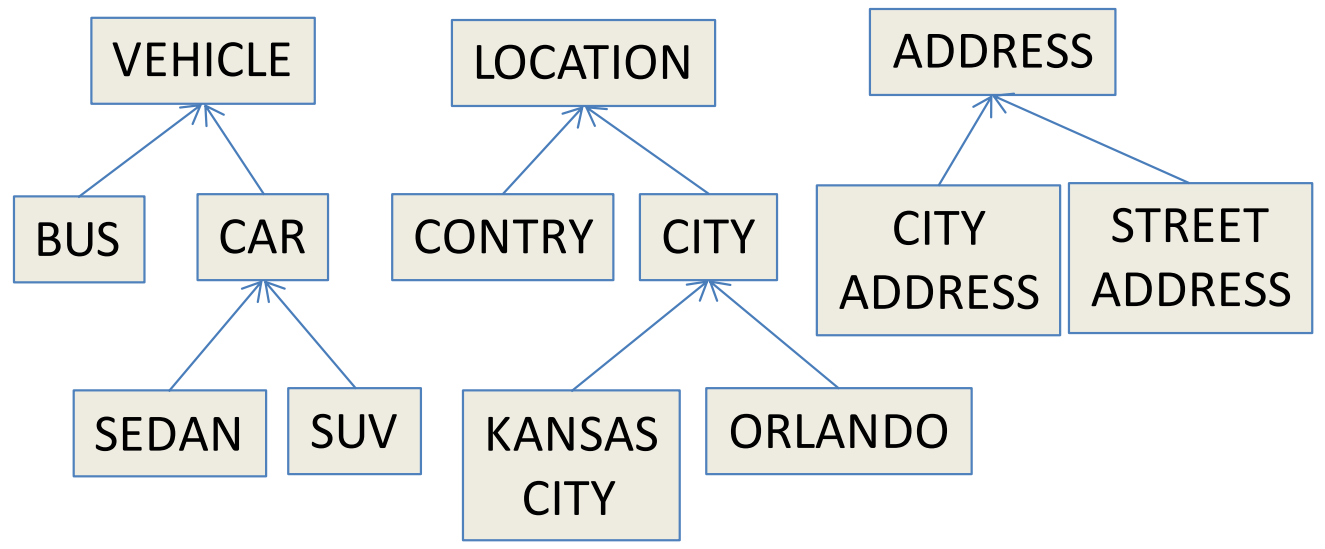

Fig. 1. Ontology of 3 taxonomies: *Vehicle, Location,* & *Address*

*city*, *address*}, $s_3.o$ = {*SUV*, *street_address*}. The domain set for this example is {*vehicle, location, address*}. Semantically, $s_1$ and $s_2$ are sibling services generating similar output under a common output abstraction {*vehicle, location*} while $s_3$ is sibling to this abstraction under a common abstraction {*vehicle*}. In other words, all the three services should belong to one categorical cluster. Suppose that the temporal online order in which each of the three services are observed into the system (i.e., the timeline over which they are first published) is $s_1 \rightarrow s_2 \rightarrow s_3$. The proof of the converged cluster space depending on the *service selection order* is given in the Appendix.

***Problem of Disjoint Category Assumption*** [5, 21-29, 40]: In most distance measure based learning approaches, the basic assumption underlying the learning problem is that clusters are disjoint. In other words, if a particular service belongs to one cluster then it cannot belong to another cluster within the same cluster space. However, this assumption is not applicable for services since a service can actually be categorized in more than one way. For an example, a car rental service having *Output* = {*car_profile, pick-up location*} can also be used as a car lookup service.

***Problems of Integrated Similarity Measure*** [18-20]: Integrated Similarity Measure based learning is a very popular approach where the measure is a linearly weighted combination (mostly a simplex) of all the functional service features (*Input*, *Output*, *Pre-condition*, *Result*) so as to produce a single distance score. However, integrated measures suffer from some serious drawbacks that are discussed below:

i) ***Effect of I/P match over O/R match***: In an integrated approach, it may happen that two service descriptions are exactly similar in terms of their *Input (I)* and *Pre-condition* (*P*) features while different in terms of their *Output (O)* and *Result (R)* features. (Note that the format definition of I, O, P, R is given in Section 4.) A high match in *I* and *P* can shift the overall similarity score beyond the chosen threshold even if there is a low match in *O* and *R*. This effect can be reduced to some extent by carefully choosing weights for each of the features [20]. However, this method also does not guarantee the elimination of this problem for all cases. For an example, in a case where the number of *Input* parameters of the two compared services is significantly higher than the number of *Output* parameters and there is an exact match between their *Input* we may see that the two services clustered together are similar. Thus, two services may be incorrectly clustered together even though their output and effect are different.

ii) ***Loss of Subsumption Match Information***: As the integrated approach provides an overall similarity score based on some similarity distance metric, it is impossible to discern from this score whether the match is a *plug-in*, *subsume*, or *vector space neighborhood match* [6]. Thus, it may be possible that a high similarity score between two services may actually be the result of semantic relatedness between the corresponding feature concepts (such as *vehicle* and *car pickup location*) instead of semantic subsumptive similarity (such as *vehicle* and *car*). The lack of subsumptive information results in loss of semantic granularity over a hierarchical cluster space. In other words, in the integrated approach we do not have the notion of a lower level cluster as a semantic subclass of the higher level cluster from which it is identified. The lack of semantic taxonomic organization substantially increases query matching computation because: (i) it is difficult to identify the optimal hierarchical level of the cluster space where the average precision and F-score can be maximized, and (ii) for a complete query match, we require an exhaustive search within each cluster since a match with the mean of a cluster (considering it to be the identity of the cluster) does not ensure that the entire cluster can be considered as a solution to the query.

iii) ***Effect of the Assumption of Task-based Query***: Integrated approaches assume user queries are formalized as desired *tasks*. Due to this reason, it is quite obvious to assume a one-to-one mapping (i.e. injection) between an atomic task and a matching service. As per injection mapping, when a query match is given both *Input* and desired *Output* components of an atomic task, it should be satisfied by the same service. This ignores a plausible scenario where a set of "*end*" services is responsible for generating the desired *Output* of an atomic task while another set of "*source*" services can take in the given *Input* of the same query such that the source services and the end services can be composed directly or indirectly (using intermediary services). Thus, integrated clustering of services may lead to a poor recall with respect to query matching if we do not distinguish between source services, intermediate services, and end services.

## IV. SEMANTIC TAXONOMICAL CLUSTERING (STC)

In this section, we formalize the problem of service clustering and propose *Semantic Taxonomical Clustering* – an alternative novel algorithm for service category learning.

### A. *Problem Statement*

In this section we formalize the problem of service clustering in the following manner: *Given a dynamic set of services* $S = \{s \mid s = \langle I, O \rangle \ni I = \{p \mid p \in \Delta\} \,\&\, O = \{q \mid q \in \Delta \,\&\, \forall q;\, \nexists p \;\exists q \equiv p$ *generate a set of "feature similar" clusters* $CS = \{C_i \mid C_i \subseteq S \,\&\, \forall s \in C_i\,;\, s_i \overset{IO}{\equiv} s_j;\, i \neq j\}$.

The underlying assumptions of the problem definition are:

- There exists a countable domain collection *D.*
- *D* is completely identified and structured into ontologies ($\Delta$).
- *D* is not *covered* (i.e., possibility for addition of new domain ontologies or new domain concepts is always open).

In the above formalism there are several observations that are noteworthy:

- All input and output parameters belong to an ontology $\Delta$. In other words they are semantically defined within an ontology. Such definition can be either dynamic (generated on-the-fly) or borrowed from existing ontologies.
- For all input parameters there cannot exist any output parameter that is *semantically equivalent* (distinct from data type equivalency). In other words, the semantic definition of parameter types has to be unique. This restriction is imposed because for services, if an output is semantically equivalent to the input, then behaviorally, the service always remains in the same functional state and can be ignored.
- *Feature Similarity* (denoted as $\overset{IO}{\equiv}$) is a stratified way of matchmaking services where services are pair-wise matched according to a single functional feature (either Input (*I*) or Output (*O*)). Details are given in Section IV. C.

The set of services is dynamic which means that new services in the given domains can be added into the set non-deterministically and existing services can be deleted non-deterministically

### B. Service Matchmaking

The fundamental operation required for service category learning in general is pair-wise *matchmaking* of service descriptions. Matchmaking is a similarity computation operation (in the context of the proposed algorithm it is called *feature similarity* and will be defined in Section IV.C) over a pair of service/query descriptions that maps a defined similarity measure function into a Real or Boolean space of a similarity score. Such matchmaking is required for clustering service registries into groups of similar services to prune the search query space.

Service matchmaking results in significant accuracy enhancement if the service/query descriptions are semantic. Semantic services are described using OWL-S [31]. The underlying mathematical foundation of most of these languages is Description Logics (DL) [32]. In such a framework service/query descriptions can be modeled as a bag of DL *concepts* that are already defined within a set of domain ontologies. The semantic service matchmaking problem then essentially becomes subsumption testing of *concepts* that have DL based definitions and are used for semantically describing services [11, 33]. Semantic service matchmaking can be of four types: (i) *exact match*, (ii) *plug-in match*, (iii) *subsume match*, and (iv) *sibling match*. In most research works as in [6 - 7, 10, 33 – 36] the first three types have been included into the service matchmaking algorithms while the fourth type has generally been neglected. Before proposing the match making algorithm (called *g-subsumption*) in this paper we first need to lay down a general background of semantic matchmaking and its four cases as follows:

***i) Exact Match*** **(EM)**: The Exact Match is a case of semantic matchmaking between two semantic descriptions where:

For every DL concept within one description, there exists a corresponding equivalent DL concept within the other description. The two descriptions (i.e. definition statements) are logically equivalent.An EM example is given in the Appendix.

ii) ***Plug-in Match*** **(PM)**: Plug-in Match is a case of semantic matchmaking between two semantic descriptions where:

There exists at least one DL concept within one of the descriptions that definitionally satisfies (i.e. subsumed by) at least one DL concept within the other description. The former description definitionally satisfies (i.e. subsumed by) the latter. An PM example is given in the Appendix.

iii) ***Subsume Match*** **(SM)**: The Subsume Match is just the inverse match of the Plug-in match. Hence, in the previous example $s_1$ has a *subsume match* with $s_3$.

iv) ***Sibling Match*** **(SBM)**: The Sibling Match is a case of semantic matchmaking between two semantic descriptions where:

1) There exists at least one DL concept within one of the descriptions that by definition satisfies (i.e. subsumed by) or has the least common subsuming concept with at least one DL concept within the other description
2) There exists at least one DL concept within the latter description that definitionally satisfies (i.e. subsumed by) or there has least common subsuming concept with at least one DL concept within the former description

Both the descriptions have a least common subsuming parent description. An SBM example is given in the Appendix.

### C. G-Subsumption Matchmaking

Most semantic matchmaking algorithms in literature have employed DL-reasoners (such as PELLET [37], FACT++ [38], etc.) for subsumption computation [10, 33 - 34]. However, DL-based subsumption reasoning can be intractable even for relatively simplistic languages within the DL family [32].

In this section, we propose a novel matchmaking algorithm, called *g-subsumption*, for computing the four cases of service matchmaking (in Section 4.C). This is an alternative non DL-reasoner based linear time algorithm that is based upon a new bit encoding technique, called *BaseOnto-encoding*. The *BaseOnto-encoding* dynamically assigns bit codes to service descriptions based on their DL based semantic definitions as explained in Section 4.C. In *g-subsumption*, a given DL based service description is partitioned into its corresponding features so as to form several conjunctive sub-descriptions (a process called *feature stratification*). In general, these features would be the four functional features – (i) *Input (I)*, (ii) *Output (O)*, (iii) *Pre-Condition (P)*, and (iv) *Result (R)*. Each of these 4 functional features is explained below:

***Input (I)***: *Input* of a service is a sub-description that includes DL concepts that are used to define the types of input parameters of the service. For an example, for the car rental service $s_1$ the input sub-description is: $hasInput.CustomerName \sqcap hasInput.CustomerID$

***Output (O)***: *Output* of a service is a sub-description that includes DL concepts that are used to define the types of output parameters of the service. For an example, for the car rental service $s_1$ the output sub-description is: $hasOutput.AutoSpecification \sqcap hasOutput.RentConfirmation$

***Pre-condition (P)***: *Pre-condition* of a service is a sub-description that defines the environment state set required to be satisfied before the service can be invoked. For an example, for $s_1$ the pre-condition can be $hasPreCondition\,.P_{s_1}$ where $P_{s_1}$ is defined as:

$$P_{s_1} \equiv [Service(s_1)] \wedge [\exists x; DOI(x) \wedge isWeekDay(x)] \wedge \begin{bmatrix} \exists y, z; Customer(y) \wedge hasAge(y, z) \wedge \\ isGreaterThan(z, 18) \end{bmatrix} \rightarrow executable(s_1)$$

The pre-condition states that if $s_1$ is an instance of the DL concept *Service* and if *x* is an instance of the DL concept *Day Of Invocation* (*DOI*) such that *x* is a week day and also if *y* is an instance of the DL concept *Customer* such that *y*'s age is greater than 18, then $s_1$ can be executed.

***Result (R)***: *Result* of a service is a sub-description that defines the new environment state and is generated by the service as a result of its execution. In the example of the car rental service $s_1$ the result can be $hasResult.\,R_{s_1}$ where $R_{s_1}$ is defined as:

$$P_{s_1} \equiv [Service(s_1)] \wedge [executed\ (s_1)] \wedge [\exists x; Car(x) \wedge hasOutput(s_1, x)] \rightarrow deductInventory(CarInventory, x)$$

The result states that if $s_1$ is an instance of the DL concept *Service* and if $s_1$ is executed and if the output instance of $s_1$ (*x*) is an instance of the DL concept *Car* then as effect *x* is deducted from the DL concept *CarInventory* representing the inventory of cars.

In our context, we include only the first two features (i.e. *I* and *O*) while an in-depth study over the other two features is left for future work. After the given service description is *feature-stratified* each of the two sub-descriptions (i.e. *I* and *O*) so formed is furthered pre-processed into a data structure called *g-array* where g = (*I*, *O*). The *g-array* groups all the object concepts within the DL sub-expressions as a set. Hence, the example car rental service $s_1$ has an *I-array = {CustomerName, CustomerID}* and an *O-array = {AutoSpecification, RentConfirmation}*. After the *feature-stratification* process is done each of the *g-arrays* are bit-codified as per the proposed *BaseOnto-encoding* algorithm (detailed in Section 4.D). Let us assume, for the sake of the current discussion, that in the above example *I-array* is encoded as *{DLcode(CustomerName), DLcode(CustomerID)} = {M, N}* where *M* and *N* are bit strings and the *O-array* is encoded as *{DLcode(AutoSpecification), DLcode(RentConfirmation)} = {X, Y}* where *X* and *Y* are bit strings. During the encoding phase, the *g-subsumption* algorithm does a global *BaseOnto-encoding* over the entire *g-array*s by ORing all the individual member *b-codes* together, and thus form two corresponding bit string *b-codes*, say $P = (M \vee N)$ and $Q = (X \vee Y)$. The global b-code is termed as *g-code*. Hence, any given DL-based service description is reduced to a *feature-stratified* set of two *g-codes*: $\{P, Q\}$.

We now define our proposed similarity measure mentioned earlier called *feature similarity (FS)* as follows:

**Definition 1**: *Feature Similarity (FS)* is a measure that is defined over the function $\sqsupset^g$ (called g-relation) that maps a pair of g-codes of two services into a 5-ary service match space (denoted g-M) = {0, 1, 2, 3, 4} where 4 represents an exact match (EM), 3 represents a plug-in match (PM), 2 represents a subsume match (SM), 1 represents a sibling match (SBM), and 0 represents no match. ■

The g-code sub-space = {1, 2, 3, 4} is called the space of feature similar g-arrays (or g-$M^{FS}$). If two given g-codes (say, $P_1$ and $P_2$ corresponding to the I-code of two services $s_1$ and $s_2$) can be mapped into I-$M^{FS}$ then $s_1$ is said to I-feature similar to $s_2$ (denoted as $s_1 \overset{I}{\equiv} s_2$).

It is to be understood that g-relation is undefined over the four algebraic operations: {+, -, *, /}. However, $\sqsupset^g$ generates an order in terms of strength of similarity where the order is defined over the sub-space g-$M^{FS}$ such as 4 > 3 > 2 > 1 > 0. The g-relation function is implemented by the g-subsumption matchmaking algorithm as shown below.

**Algorithm:** *g-subsumption*
**INPUT**: $s_1$.gA, $s_2$.gA // *g-arrays* of $s_1$ and $s_2$
**OUTPUT**: {0, 1, 2, 3, 4} // the *g-M* space

```
START
s1.g = s2.g = 0
/* checkDomain checks whether a given g-array is already defined in
the set of domain ontologies */
if (checkDomain(s1.gA) == false){
// dynamically assigns a unique b-code to s1
    s1.g = BOEncode(s1.gA) //BO-Encode: BaseOntoEncoding (V-C)
}else{
// extracts the already existing b-code for s1
    s1.g = getDLCode(s1.gA)}
if (checkDomain(s2.gA) == false){
// dynamically assigns a unique b-code to s2
    s2.g = BOEncode(s2.gA)
}else {
// extracts the already existing b-code for s2
    s2.g = getDLCode(s2.gA)
}
if (s1.g ∨ s2.gA == s1.gA){
  if(s1.g ∨ s2.gA == s2.gA)
     FS_strength = 4 // case of exact match
  else FS_strength = 3 // case of plugin match
}else if(s1.gA ∨ s2.g A== s2.gA)
  FS_strength = 2 // case of subsume match
else if(s1.gA ∧ s2.gA != 0){ // case of sibling match
  FS_strength = 1
  Abstract_Parent.gA = s1.gA ∧ s2.gA
}
Return  FS_strength
END
```

## D. BaseOnto-encoding

For a given SOA based system we assume that there exists a terminology $\Delta^c$ consisting of *concepts*. These concepts may be ordered partially according to the subsumption relation $\sqsubseteq$. Hence, a corresponding *concept taxonomy* ($T^c$) can be formed out of $\Delta^p$. In a similar fashion we also assume the existence of a terminology *R* consisting of *roles* (relations) from which a corresponding *role taxonomy* ($T^r$) can be formed. Thus, the *domain space* includes both $T^c$ and $T^r$.

In order to encode the *domain space* we first include the universal concept ⊤ and an empty concept ⊥ within a given *domain space* to transform the *domain space* into a lattice structure where the order relation is the subsumption relation $\sqsubseteq$. The lattice so formed can be seen as a directed acyclic graph, called *domain ontology* (denoted as $T^{\Delta}$), with a single root concept ⊺ and a single leaf concept ⊥. The root concept is called the *universal parent* (since it subsumes any DL concept) while the leaf concept is called the *universal child* (since it can be subsumed by any DL concept). We define a *parent* and a *child* concept as follows:

**Definition 2**: A *parent concept* $c_i$ is a concept within the *base ontology* (i.e. initially given ontology) such that there exists at least one concept $c_j$ within the same *base ontology* such that $c_i \sqsubset c_j$. $c_i$ is said to be the parent of $c_j$. ■

**Definition 3**: A *child concept* $c_i$ is a concept within the *base ontology* such that there exists at least one concept $c_j$ within the same *base ontology* such that $c_i \sqsupset c_j$. $c_i$ is said to be the child of $c_j$. ■

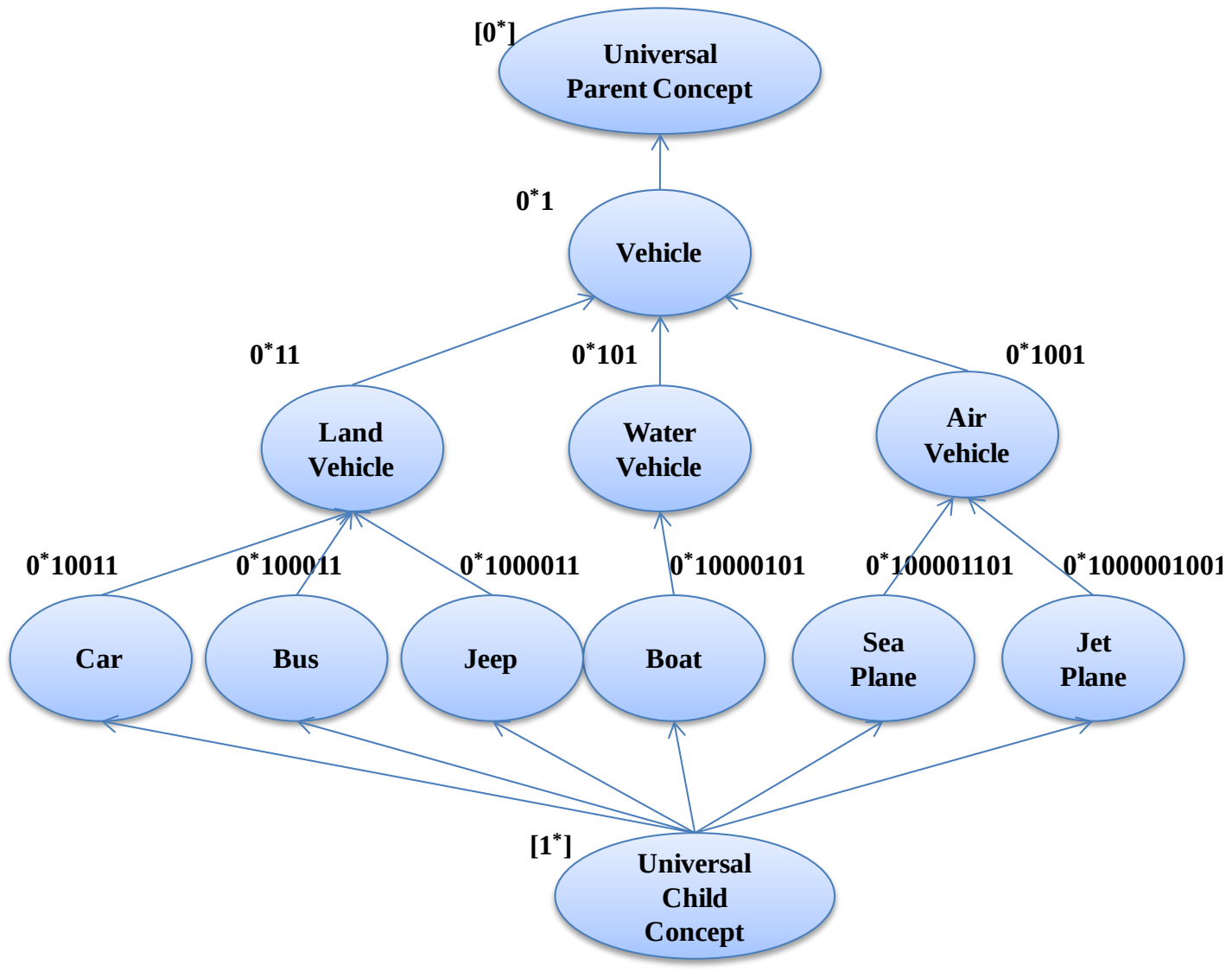

Fig. 2. A *Vehicle* Base Ontology (Encoded by BaseOnto-encoding)

The *BaseOnto-encoding* algorithm is a simple topological spanning over the corresponding graph to define a base ontology $T^{BS}$ starting with assigning code [$0^*$] to the universal parent ⊺ and finishing with assigning the code [$1^*$] to the universal child ⊥. The superscript [$^*$] means that the 0/1 is repeated over an *n* bit string length where *n* refers to the current number of concepts within the *domain space*. During the topological spanning a new 1 bit, called the *most significant bit*, is assigned to a concept that signifies its unique identity. The assignment is done at the code string position that corresponds to the order of the topological span. For an example, in Fig. 2, the order of the visit to the concept *Car* is 5. Hence, a 1 bit is assigned to the $5^{th}$ position of the code string. While assigning code to a concept at a particular visit all the codes of its parent concepts are ORed together so that all of their respective uniqueness can be inherited. This ORed code is then concatenated together with the newly assigned 1 bit. Thus, the concept *Car* in Fig. 2 is encoded as $0^*10011$ where the 1 bit at position 1 is inherited from the code of its sole parent *LandVehicle* (whose code is $0^*11$). Codes assigned in this manner are called *b-code*s. The BaseOnto-encoding algorithm is outlined in Appendix.

A very important property of the encoding algorithm is that it always assigns a unique code to any given concept within the *base ontology*. The uniqueness is guaranteed by the assignment of the most significant bit during the topological span. Once the *base ontology* is encoded, we can very efficiently compute whether two given base concepts are mutually subsumptive by using the following theorem:

**Theorem 1: *Subsumption Test Validity*.** $c_x \sqsubseteq c_y$ iff [*b-code*($c_x$) ∨ *b-code*($c_y$) = *b-code*($c_x$)] where $c_x, c_y \in T^{BS}$ .

***Proof***: If $c_x \sqsubseteq c_y$ then $c_x$ must inherit all the 1 bits of $c_y$ according to the *BaseOntoEncoding* algorithm. Hence, [*b-code*($c_x$) ∨ *b-code*($c_y$)] contains all the common inherited 1 bits of $c_y$ and $c_x$. For the non-inherited 1 bits of $c_x$, there can only be corresponding 0-bits of $c_y$ since all the 1-bits of $c_y$ have already been ORed up in [*b-code*($c_x$) ∨ *b-code*($c_y$)]. Thus, [*b-code*($c_x$) ∨ *b-code*($c_y$)] will also contain all the non-inherited 1 bits of $c_x$. This implies that if $c_x \sqsubseteq c_y$ then *b-code*($c_x$) ∨ *b-code*($c_y$) = *b-code*($c_x$).

If [*b-code*($c_x$) ∨ *b-code*($c_y$) = *b-code*($c_x$)] then all the 1 bits of $c_x$ are preserved in the result. Now if $c_y$ is not identical with $c_x$ (i.e. $b - code(c_y) \neq b - code(c_x)$) then there can be two cases: (i) *b-code*($c_x$) must contain a set of 1 bits that are not contained in *b-code*($c_y$), and (ii) *b-code*($c_y$) must contain a set of 1 bits that are not contained in *b-code*($c_x$). The second case is a contradiction to the initial assumption that [*b-code*($c_x$) ∨*b-code*($c_y$) = *b-code*($c_x$)] since the result after ORing will be *b-code*($c_y$) instead of *b-code*($c_x$). Hence, the first case is true. Since all the 1 bits of *b-code*($c_y$) are common to *b-code*($c_x$) it therefore implies that $c_x \sqsubseteq c_y$. ■

The above theorem proves that *g-relation* function ($\sqsupset^g$) is sound and complete over the base ontology $T^{BS}$. The time complexity of *g-relation* over $T^{BS}$ is Θ(N/W) where N is the total number of concepts in $T^{BS}$ and W is the word length of a particular computational model (e.g., 64 bits computer). An example for *b-code subsumption* can be that of the concept *Car* and the concept *LandVehicle* in Fig. 2. *Car* ⊑ *LandVehicle* = $0^*10011$ ∨ $0^*11$ = $0^*10011$ = *Car*. Thus, the concept *LandVehicle* subsumes the concept *Car*.

### *E. Taxonomical Clustering Spaces*

In the proposed *STC* algorithm, we define a cluster space as a set of *service taxonomies*. We first formally define a service taxonomy as follows:

**Definition 4**: A g-type *service taxonomy* (denoted as $T^g$) is a partial-order $\langle s, \sqsupset^g \rangle$ where *s* is a service and the order is the *g-relation* $\sqsupset^g$ where *g* = *{I, O}* s.t. there exists a unique supremum (or *least specific predecessor*) called the *root service*. ■

*Service taxonomy* (in brief *taxonomy*) $T^g$ has some basic properties as discussed below:

- A taxonomy is a cluster of *feature similar* (*FS*) services where the feature set *g* = *{I, O}*.
- *Feature similarity* in a service taxonomy can be of four types: (i) *exact*, (ii) *plug-in*, (iii) *subsume*, and (iv) *sibling* (as discussed in the previous section).
- Taxonomy is a stratified cluster of feature similar services. This is because the *g-subsumption* relation is either with

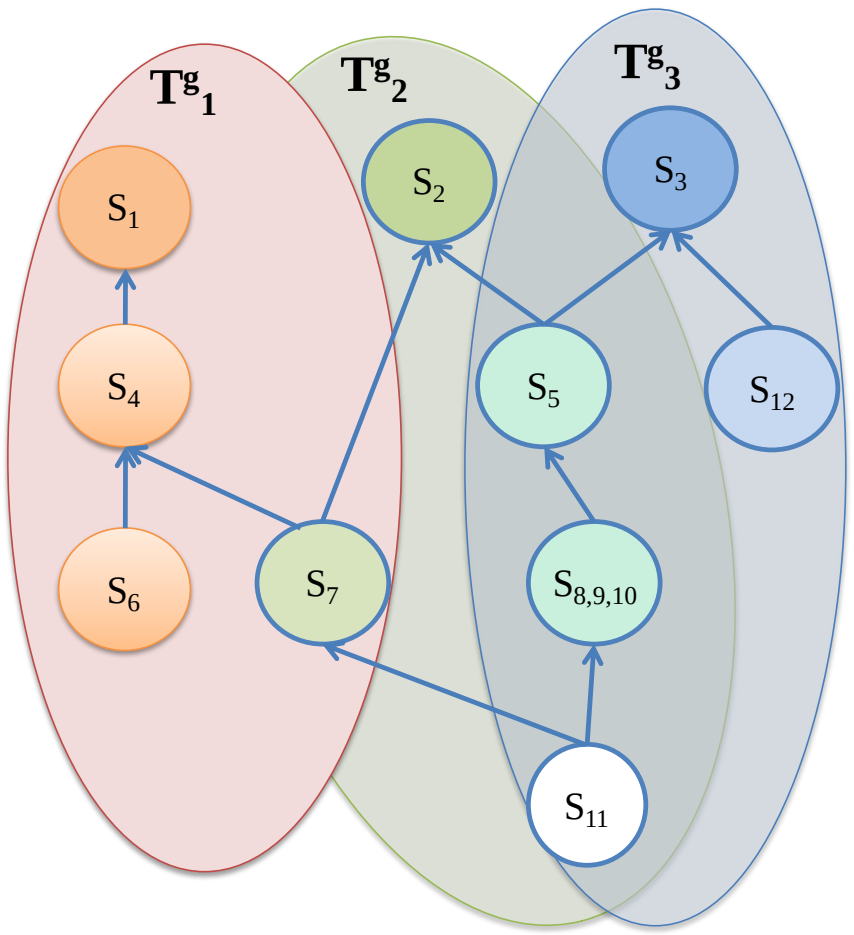

Fig. 3. g-Taxonomical Cluster Space

respect to *Input* (*I*) or *Output* (*O*).

- *Feature similarity* with respect to a taxonomy is non-distance based. In other words, the similarity condition is not based on any measure but rather on the type of semantic subsumption match (*exact/plug-in/subsume/sibling*).

We now define a taxonomical cluster space as follows:

**Definition 5**: A *g-type taxonomical cluster space* (denoted as *CS-T^g*) w.r.t to a particular functional feature *g* is a dynamic set of *g-type taxonomic clusters*. ■

*g-type Taxonomical cluster space* (in brief, *g-cluster space*; Fig. 3) has several properties that make it unique from cluster spaces generated in conventional learning algorithms. They are discussed below:

- A taxonomical cluster space is dynamic (i.e. addition and deletion of member services within taxonomies can be non-deterministic). Hence, as we will see in the proposed STC algorithm, the cluster space supports online learning.
- There are two types of cluster spaces possible: (i) *O-cluster space* (where the feature similarity is over the *Output* feature) and (ii) *I-cluster space* (where the feature similarity is over the *Input* feature).
- The member taxonomies within the cluster space are not necessarily disjoint from each other. This is because a particular service can have a *g-subsumption* plug-in match with more than one service, each of which is a member of separate taxonomies. For an example, a car rental service having *Output = {car info, rental confirmation}* may have *plug-in* matches with both a vehicle rental service having *Output = {vehicle confirmation}* and a vehicle lookup service having *Output = {vehicle info}* with the *O-cluster space*. In this example the vehicle rental service and the vehicle lookup service belong to two different taxonomies (i.e. taxonomies having two distinct root services).
- The converged topology of the cluster space is independent of the *order of sample selection*. In other words, the temporal order in which services are published into the system does not affect the final cluster space topology (i.e. the number of taxonomies and the partial ordering within each of the taxonomies). This will be evident once we introduce the taxonomical clustering algorithm in the next section.

We now define an *MSP* and an *LSC* of a particular selected service *s* below. These two structures form the basis of the STC algorithm that will be described in the next section.

**Definition 6**: *MSP* (or *Most Specific Parents*) of a given service $s$ is a set of services such that: $\forall p \in MSP\ ;\ p \sqsupset^{g} s \wedge \nexists q \ni p \sqsupset^{g} q \sqsupset^{g} s$. ■

**Definition 7**: *LSC* (*Least Specific Children*) of a given service *s* is a set of services such that: $\forall m \in LSC\ ;\ s \sqsupset^{g} m \wedge \nexists n \ni s \sqsupset^{g} n \sqsupset^{g} m$ . ■

### F. STC Clustering Algorithms

The basic outline of the proposed *STC* learning algorithm involves "*semantically inserting*" a randomly selected service from a dynamic service set into one or more *g-taxonomies* within the corresponding *g-cluster space*. The insertion of random service is done by searching for the *most specific parents* (*MSP*) and the *least specific children (LSC)* of the service (Fig. 4). The algorithm utilizes an important property of a *g-taxonomy* to improve the clustering efficiency. The property has been given in the form of a theorem below:

**Theorem 2**: If for a selected service *s* there exists a non-empty MSP and if there exists a non-empty LSC of *s* then $\forall p \in LSC\ ; \exists m \in MSP \ni m \sqsupset^{g} p$ .

***Proof***: As the selected service $s \sqsupset^{g} p$ and as the MSP exists hence, $\exists m \ni m \sqsupset^{g} s \sqsupset^{g} p$. ■

The implication of the above theorem is that for semantically inserting a selected service into a taxonomy we need to identify the MSP of the service. Once that is done then we can restrict the search for LSC of the selected service to the LSC of each member service within the MSP. This significantly reduces the search space for finding the correct taxonomic position of the selected service. If the MSP of the selected service is empty and the selected service does not have a *sibling match* with any of the existing *root services* then the selected service becomes a *root service*. In that case, the LSC of the selected service has to be identified from the entire existing cluster space. Otherwise, if the selected service has a *sibling match* then a new *abstract service* is created that subsumes the sibling services. This operation is very significant in the process of service discovery. Another implication of the above theorem is that the member services in the MSP may not belong to the same taxonomy. In other words, there may exist more than one root services $s_r$ such that $s_r \sqsupset^{g} m;\ m \in MSP$. Hence, the selected service may belong to multiple taxonomies (Fig. 4).

The *STC* algorithm returns an instantiated cluster space (CS) when given the dynamic service set *S*. This main algorithm requires two functions: (a) *findMSP* for computing the MSP of a particular service, and (b) *findLSC* for computing the LSC of a particular service. For pair-wise service matching, the g-*subsumption* algorithm is used. It returns 0 if there is no match, 1 if the argument services are *sibling match* under a common abstract parent service, 2 if the first argument service subsumes the second argument service (*subsume match*), 3 if the first argument service is subsumed by the second argument service (*plug-in match*) and 4 if the first argument service is semantically equivalent to the second argument service (*exact match*).

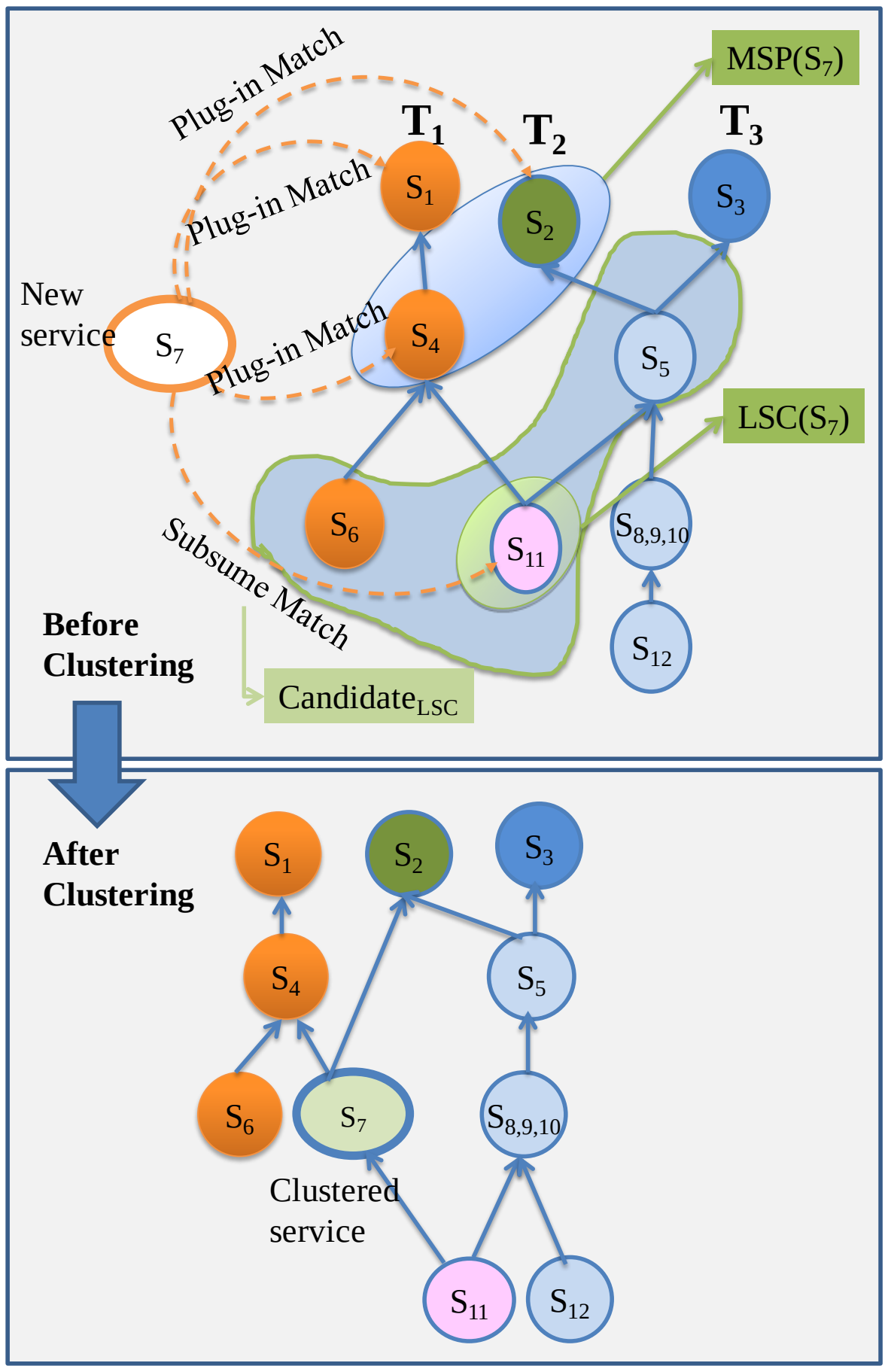

Fig. 4. STC Algorithm Outline

We now provide the mathematical proofs for the soundness and completeness of *STC* as follows:

**Theorem 3: STC Soundness**. Given a newly observed service $s$ STC clusters $s$ into a sub cluster space $N$ so that all the services in $N$ have a relationship (*g-relation)* with $s$.

**Proof***:* For any arbitrary sample space $S$ if an arbitrary sample $s$ is selected then it has to either match with one or more of the existing taxonomies (say $T_E$) or none. All possible answers in the algorithm can be represented in the form: $N_i \sqsupset^g s \sqsupset^g N_j \; where \; N_i \subseteq \{\varepsilon, n_{1i}, n_{2i}, \dots, n_{mi}\}$; $N_j \subseteq \{\varepsilon, n_{1j}, n_{2j}, \dots, n_{kj}\}$ s.t. $n_i, n_j$ are services in the sub cluster space $N$ ($N = N_i \cup N_j$ ) having g-relation ($\sqsupset^g$) with $s$. To prove the soundness we need to prove that $\left[\nexists n_p \in N_j \ni s \,!\sqsupset^g n_p\right] \wedge \left[\nexists n_q \in N_i \ni n_q ! \sqsupset^g s\right]$.

For the initial case when $N_i, N_j = \{\varepsilon\}$, $[\nexists n_p \in N_j \ni$ $s \not\sqsupset g np \wedge \nexists nq \in Ni \ni nq \not\sqsupset gs$ is trivially true.

Assuming $\exists n_p \in N_j \ni s\,! \sqsupset^g n_p$ is true. According to STC, $\exists n_r \in LSC(s) \; iff \; N_j \neq \{\varepsilon\}. \therefore n_r \in N_j$. Since $n_p \in N_j$ therefore, we can choose $n_r \ni n_r \sqsupset^g n_p$. As $\sqsupset^g$ is transitive, $s \sqsupset^g n_p$. This is contradictory to the previous assumption. Therefore, $n_p \notin N_j, \Rightarrow \nexists n_p \in N_j \ni n_p! \sqsupset^g s, \Rightarrow N_j$ is correct.

Similarly assuming $\exists n_q \in N_i \ni n_q! \sqsupset^g s$ is true. According to STC, $\exists n'_r \in MSP(s) \; iff \; N_i \neq \{\varepsilon\}. \therefore n'_r \in N_i$. Since $n_q \in N_i$ we can therefore choose $n'_r \ni n_q \sqsupset^g n'_r$. As $\sqsupset^g$ is transitive, therefore $n_q \sqsupset^g s$. This is contradictory to the previous assumption. Therefore, $n_q \notin N_i, \Rightarrow \nexists n_q \in N_i \ni n_q! \sqsupset^g s, \Rightarrow N_i$ is correct. ■

**Theorem 4**: **STC Completeness**. Given a newly observed service $s$ STC clusters $s$ into a sub cluster space $N$ such that $N$ does not falsely exclude any service that has *g-relation* with $s$.

**Proof**: To prove the completeness we need to prove that $\left[\nexists n_p \notin N_j \ni s \sqsupset^g n_p\right] \wedge \left[\nexists n_q \notin N_i \ni n_q \sqsupset^g s\right]$.

For the initial case when $N_i, N_j = \{\varepsilon\}, [\nexists n_p \notin N_j \ni$ $s \sqsupset g np \wedge \nexists nq \notin Ni \ni nq \sqsupset gs$ is trivially true.

Assuming $\exists n_p \notin N_j \ni s \sqsupset^g n_p$. According to STC, $\exists n_r \in LSC(s) \; iff \; N_j \neq \{\varepsilon\}. \therefore n_r \in N_j$. Since $s \sqsupset^g n_p$ and $s \sqsupset^g n_r$ we can choose an $n_r$ such that $n_r \sqsupset^g n_p$. Therefore, $n_p \in N_j$ (contradictory to assumption). Hence, $\nexists n_p \notin N_j \ni s \sqsupset^g n_p$. Using a similar argument for MSP(s), we can prove that $\nexists n_q \notin N_i \ni n_q \sqsupset^g s$. ■

We now provide a detailed outline of the *STC* algorithm along with the *findMSP* and *findLSC* functions that are called within *STC.*

**ALGORITHM: *Semantic Taxonomic Clustering (STC)***
**INPUT:** sample space S = {$s_1$, $s_2$, $s_3$ ….. $s_n$}, n
**OUTPUT:** cluster space $CS_{1\ldots n}$

```
START
CS = NULL // initially CS is set as empty
for count = 1 to n {
    sample = randomSelect(S);
    S = S – {sample};
    MSP = NULL;
    root = extractRootNode(CS);
   // CASE 1-A: When CS is empty (initial state)
    if(root = NULL) {
        CS = {sample};
        return CS; }
    for c = 1 to root.size{
        root[c].visited = false;
        MSP = MSP ∪ findMSP(sample, root[c]);
        for i = 1 to MSP.size {  // PLSC: Potential LSC
          PLSC = PLSC ∪ findLSC(memberOf(MSP))}
        c++; }
   findLSC(sample, PLSC, CS) }
return CS
END
```

**ALGORITHM**: ***findMSP***
**INPUT**: sample, node
**OUTPUT**: MSP of sample

```
START
sample.visited = false;
 if (node.visited = false){
   // CASE 1-B: NO MATCH
    if (g-SubsumptionMatch(sample, node) == 0) {
     MSP = NULL;
     return MSP;}
    // CASE 1-C: PLUG-IN MATCH
    else if (g-SubsumptionMatch(sample, node) == 3) {
       node.visited = true; //  this node won't be selected again
       if (node.childrensize != 0) {
         for c1 = 1 to node.childrensize {
          node = node.child[c1];
```

```
            c1++;
            findMSP(sample, node); } }
          else {
             sample.parent[sample.parentsize + 1] = node;
             sample.parentsize++;
             node.child[node.childrensize + 1] = sample;
             node.childrensize++; }}
    // CASE 1-D: EXACT MATCH
    else if (g-SubsumptionMatch(sample, node) == 4) {
        node.visited = true; //  this node won't be selected again
        sample.parent = sample.parent ∪ node.parent}
    // CASE 1-E: SIBLING MATCH
    else if (g-SubsumptionMatch(sample, node) == 1) {
        node.visited = true; //  this node won't be selected again
        sample.parent = sample.parent ∪ sample.abstractparent; }
     MSP = sample.parent;
    return MSP; }
else return MSP; }
END
```

```
==========================================
ALGORITHM: findLSC
INPUT: sample, member of sample MSP, CS
OUTPUT: LSC of sample for the member of sample MSP
==========================================
START
// CASE 2-A: When sample has a non-empty MSP
if (member != NULL) {
        for c = 1 to member.childrensize {
            if (g-SubsumptionMatch(sample, member.child[c])== 2) {
             LSC = LSC ∪ member.child[c];} }
         insertMember(CS, sample);
         return LSC; }
// CASE 2-B: When sample has an empty MSP
Else {
      for count = 1 to CS.size {
         node = extractMember(CS);
         if  (node.visited == false) {
              node.visited = true;
              if (g-SubsumptionMatch(sample, node) == 2){
                     LSC = LSC ∪ node; }
             else continue;}
         else continue; }
       insertMember(CS, sample);
       return LSC;}
END
```

## V. Results

### A. Experimental Setup

We tested the runtime performance of the proposed *STC* algorithm on a system having a CPU cycle of 1.4 GHz and a RAM of 2 GB. The performances were measured based on: (a) randomly generated synthetic services of size 1500 and (b) OWL-S TC v.2 test set of 871 web services collected from different web service registries.

For generating the random sample space we designed a simulation platform where sample spaces of different sizes (50 to 1500 web services) were randomly generated using a domain space that consisted of 10 domain ontologies (with an average number of concepts set to 300). An average parameter size of 5 was set for the simulation. Service parameters where chosen randomly from the 10 domain ontologies such that the *Input* feature of each service is distinct from the *Output* feature.

### B. Runtime Performance

The clustering performance is evaluated on the basis of: (i) average runtime for clustering under an online learning framework and (ii) effect of stratification in the proposed *STC* algorithm (as compared to a non-stratified nearest neighbor based online learning algorithm). As the learning framework is online, for both the synthetic dataset as well as the OWL-S TC dataset were drawn from the set in random temporal sequence.

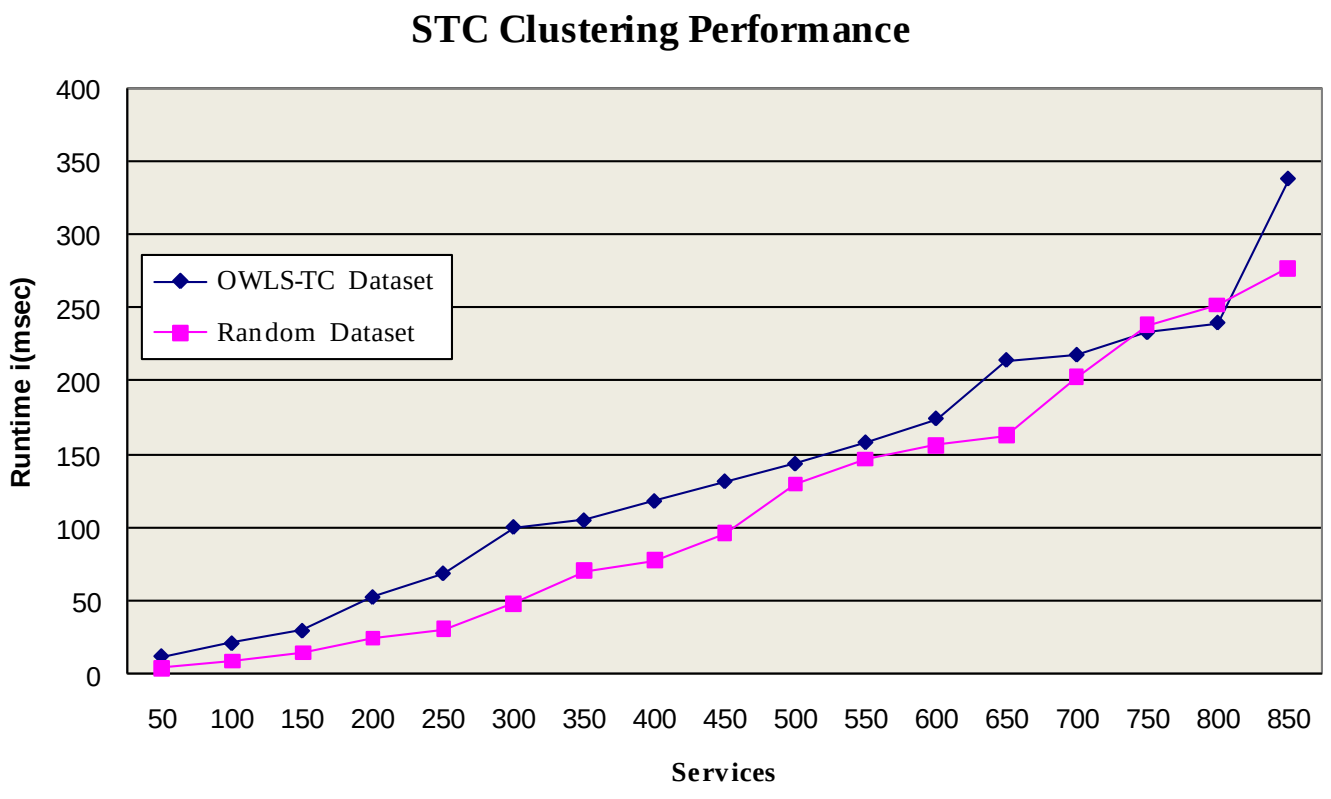

Fig. 5. Average Runtime Performance of STC

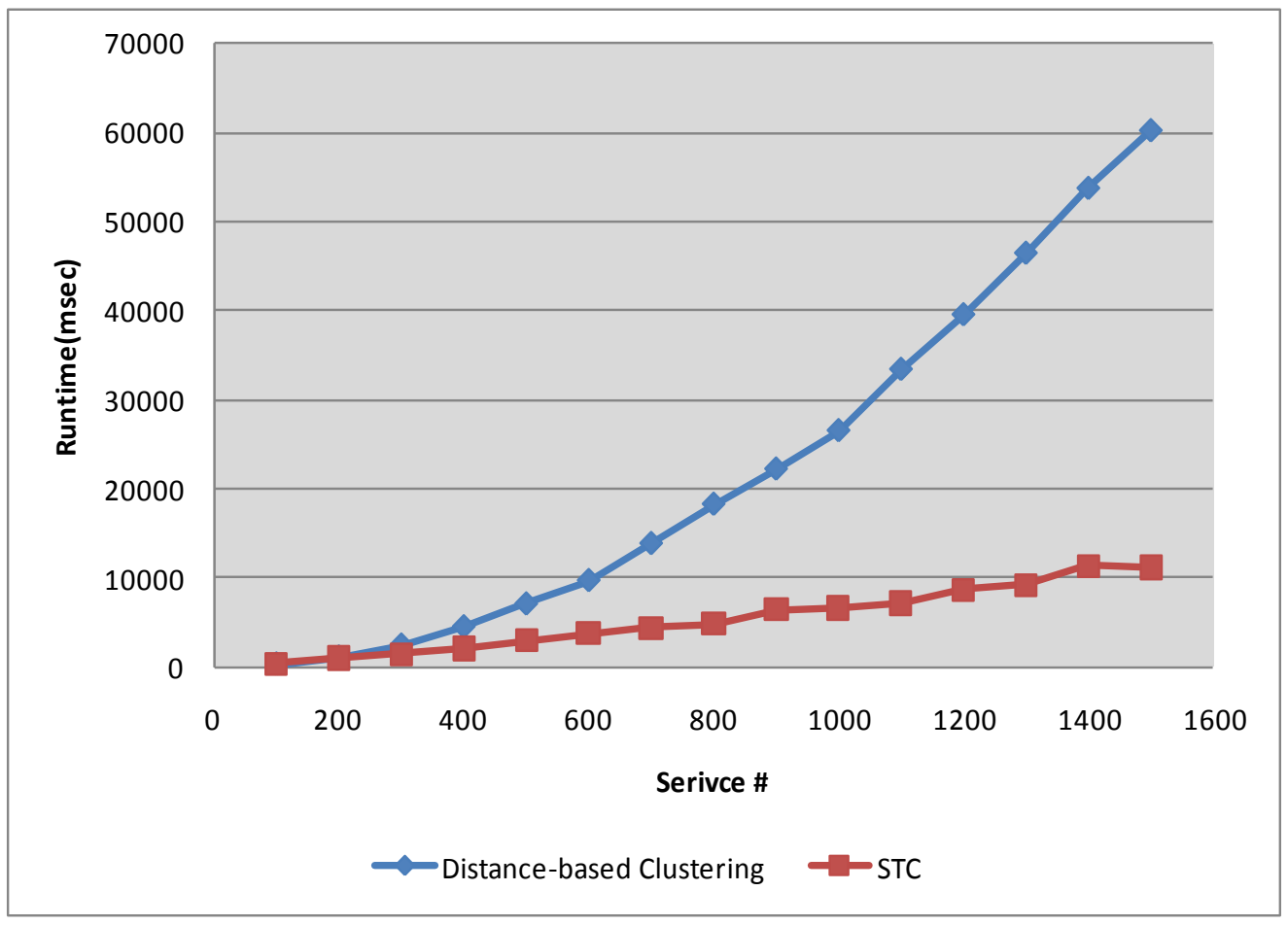

Fig. 6. STC algorithm vs. Distance based Nearest Neighbor Clustering

***Average Clustering Runtime***: When we conducted our experiment with synthetic simulated service sets we found a substantial runtime clustering performance within an average range of 0.003 seconds (for 50 samples) to 0.277 seconds (for 850 samples out of 1500 samples) (Fig. 5). For OWL-S TC set, the runtime was similar with an average range of 0.011 seconds (for 50 samples) to 0.337 seconds (for 871 samples) (Fig. 5).

***Effect of Stratified Clustering***: We wanted to observe the runtime performance improvement of the stratified clustering approach as compared to an integrated distance-based clustering approach. For that, we chose the *semantic distance measure* proposed by us in [42]. The learning algorithm that was implemented in this work was nearest-neighbor based. For the comparison we used the synthetic dataset of 1500 services. We observed a significant improvement in performance as the number of services increased (Fig. 6). This is because of two major reasons: (i) the pair-wise similarity computation for the integrated *SGPS measure* is significantly higher than that of *g-subsumption* comparison, (ii) the amortized number of pair-wise *g*-subsumption comparisons needed for the STC algorithm is significantly lower than the amortized number of pair-wise comparisons needed for the nearest-neighbor based algorithm. For evaluating the accuracy of the proposed STC algorithm we used the standard test dataset OWL-S TC v2.

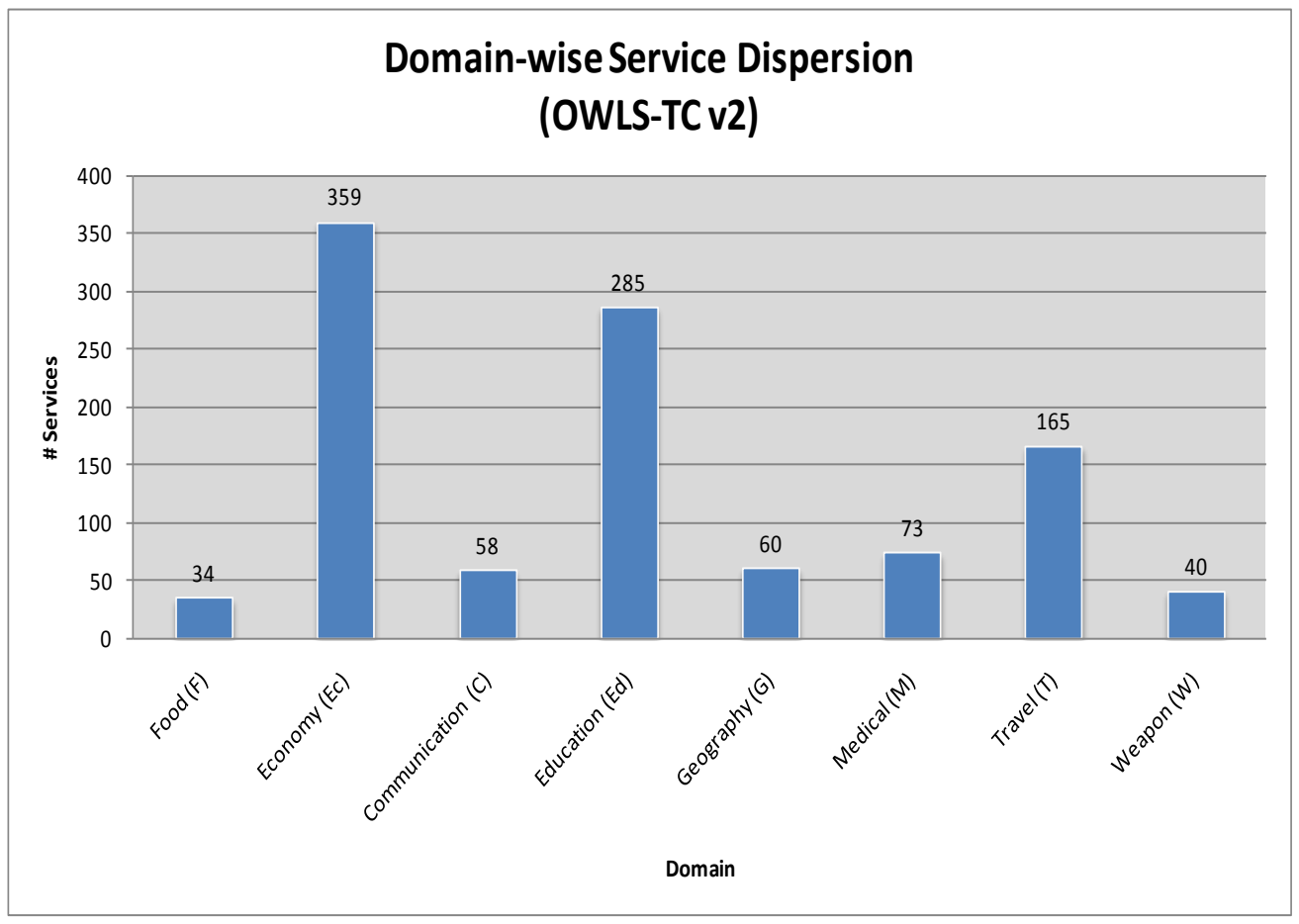

Fig. 7. Distribution of OWLS-TC v2 web services according to 8 domains

## C. STC Clustering Accuracy

***Domain-based Accuracy Measuring:*** The first objective of our experiment was to understand how close STC fits the given service descriptions as compared to the given categorization of these service descriptions in terms of their corresponding domains. In order to meet the objective, we first defined two important measures:

**Definition 8**: *Domain-Precision* with respect to a given service category *C* and a given service domain *D* (denoted as *Pr(C, D)*) is defined as the ratio of the number of services in *D* that are categorized as members in *C* (say $N_{C,D}$) by any given learning algorithm *L* vs. the number of services categorized in *C* (say $N_C$) by *L*. Numerically this means: $Pr(C, D) = N_{C,D} / N_C$ . ■

**Definition 9**: *Domain-Recall* with respect to a given service category *C* and a given service domain *D* (denoted as *Re(C,D)*) is defined as the ratio of the number of services in *D* that are categorized as members in *C* (say $N_{C,D}$) by any given learning algorithm *L* vs. the number of truly correct services in *D* (say $N_D$). Numerically this means: $Re(C, D) = N_{C,D} / N_D$. ■

Note that the above definitions give us a way of computing precision and recall of a given learning algorithm that is completely independent of any query (as opposed to the more conventional query based precision/recall computation that we will discuss later in this section). However, the definition is still a subjective evaluation as it requires human judgment for estimating $N_D$ (number of truly correct services in domain *D*).

While evaluating the *domain-precision* and *domain-recall* of STC we took the standardized human evaluation included within the OWLS-TC v2 dataset. In the OWLS-TC dataset there are 8 web service domains as shown in Fig. 7. The innate assumption implied within the expert evaluated classification is that all domains are mutually disjoint. We first observed the accuracy over the *O-cluster space* of STC. We observe that there are 50 clusters that are generated (Fig. 8).

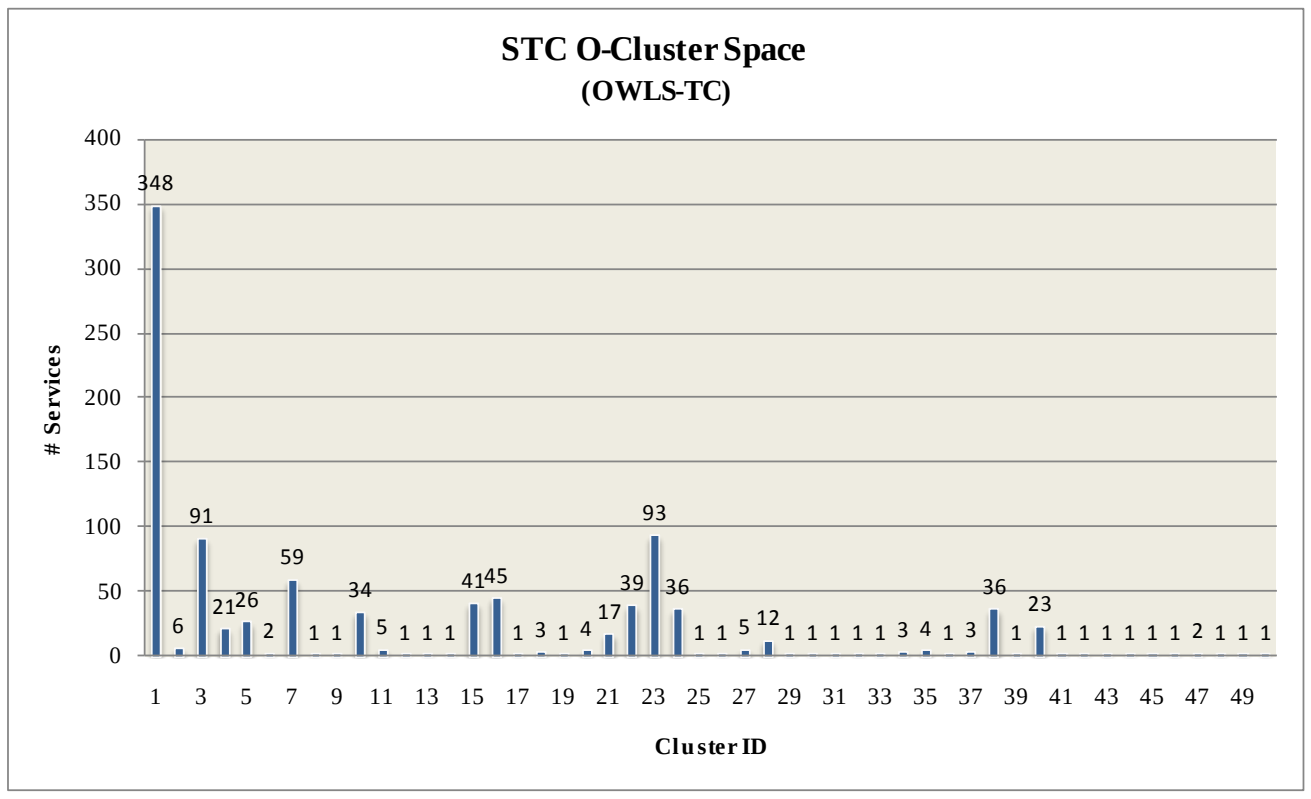

Fig. 8. Output Cluster Space generated by STC

However, out of these 50 clusters there are only 15 clusters that are significant in terms of number of services per cluster. In Fig. 9 we make a comparative analysis of the *average domain-precision* (in light blue) and *average domain-recall* (in deep blue) of each of these 15 clusters when compared to the domain-wise classified web services in the OWLS-TC dataset. The *average domain-precision* in this context is computed as: $\Pr\left(C_i^g\right) = \frac{\sum_{i=1}^{8} \Pr\left(C_i^g, D_i\right)}{8}$. Similarly the *average domain-recall* is also computed as $\mathrm{Re}\left(C_i^g\right) = \frac{\sum_{i=1}^{8} \mathrm{Re}\left(C_i^g, D_i\right)}{8}$. We observed that the average precision for almost all the significant clusters (except for one) is close to 1 while the average recall is comparatively low in most cases. Upon analysis we understand that the Output-cluster space is strong enough to represent each of the domains in OWLS-TC that supports our argument that the *Output* parameter is the most significant service feature in understanding service functionality. Hence, STC was able to reduce false positives within the Output-cluster space by restricting inclusion of services to only those that have mutual *Output space g-subsumption* matches.

A very interesting observation was made with respect to this result: most of the domains were split over the cluster space.

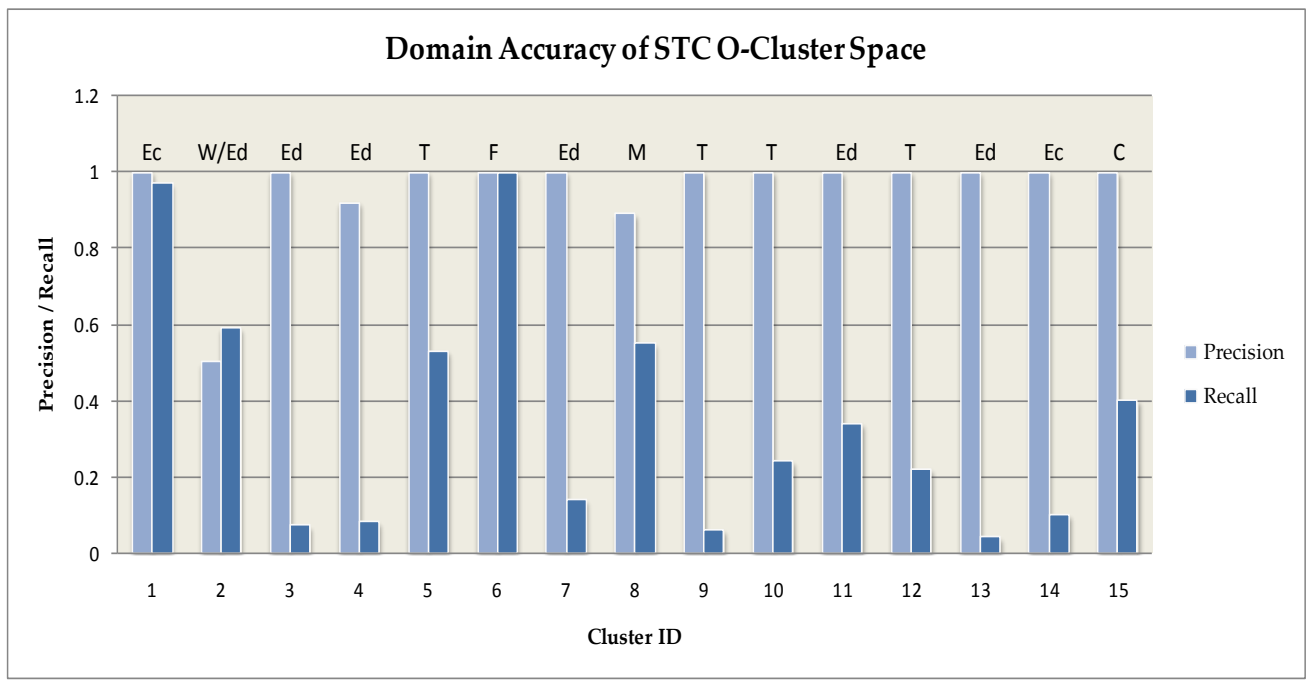

Fig. 9. Domain-Accuracy of Output-cluster space generated by STC

For an example, the domain *Economy* had been split into 2 clusters, each having average precision 1 while one having the recall significantly higher than the other. This phenomenon occurs because STC did not assume the clusters to be disjoint. Hence, there may be two different functionalities that describe only services within the *Economy* domain in OWLS-TC. Each functionality represents a separate (although overlapped) *Output-cluster space*.

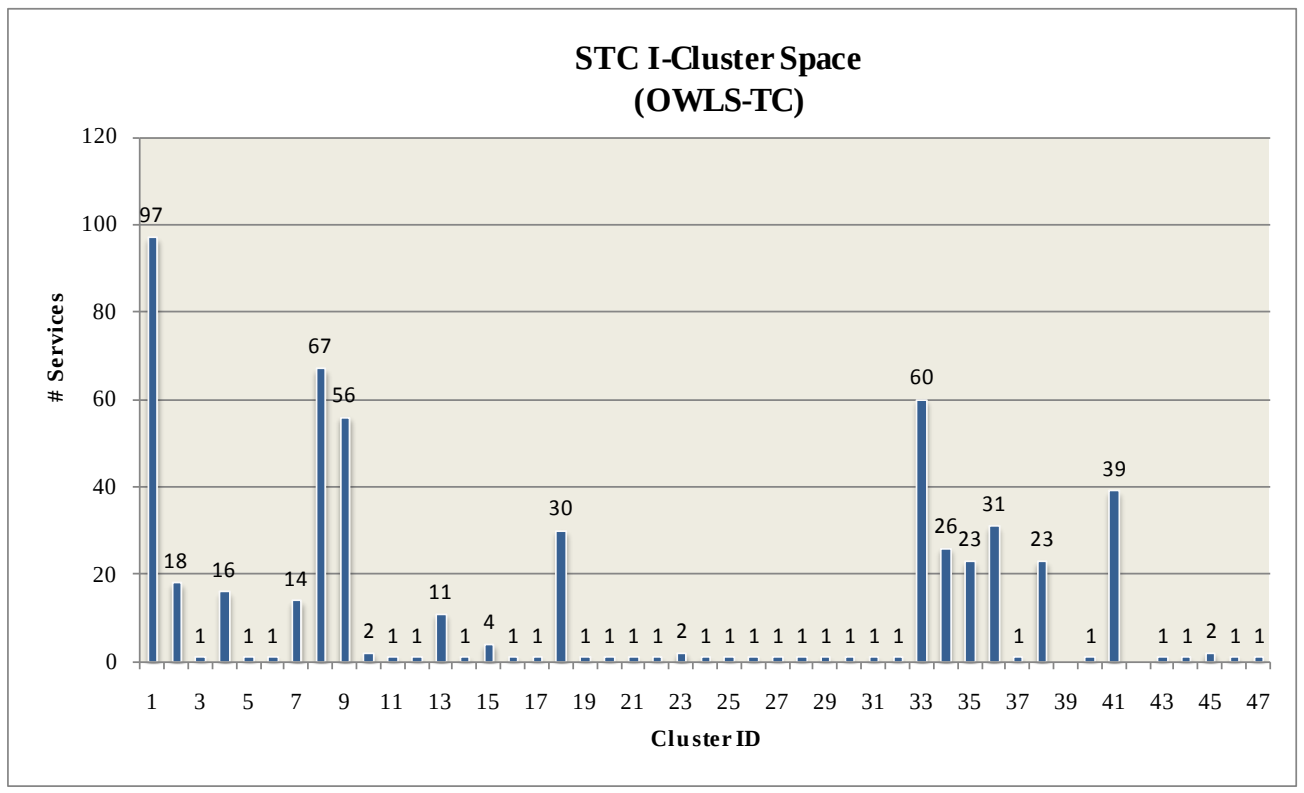

Fig. 10. Input Cluster Space generated by STC

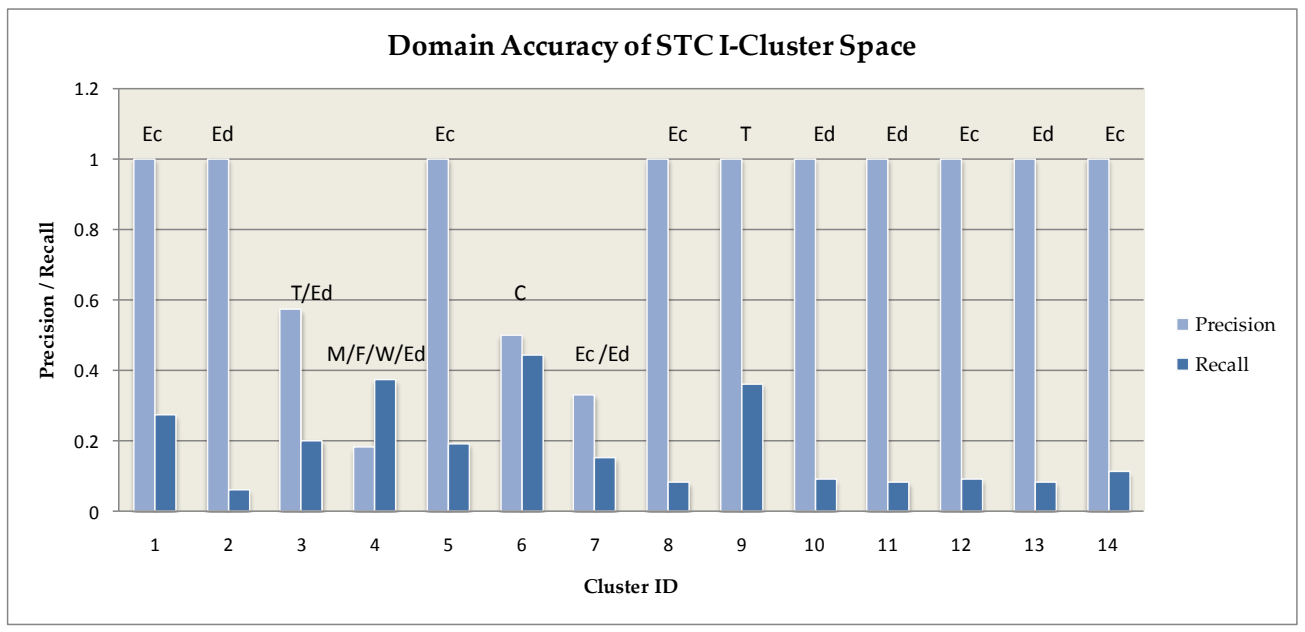

Fig. 11. Domain-Accuracy of Input Cluster Space generated by STC

When we compare this result to the *I-cluster space* generated by STC over the same dataset we find that there are 45 clusters (Fig. 10) out of which only 14 clusters are significant. On observation of accuracy in terms of *domain-precision* and *domain-recall* we saw that the overall precision and recall dropped (Fig. 11). Another very interesting observation can be made within the *I-cluster space*: majority of clusters is "*mixed bag*" in nature in the sense that more than one domain is significantly represented (in terms of precision and recall) by each of these *mixed bag* clusters. This phenomenon occurs because the *Input* feature of a service does not adequately represent its functionality. Hence, there can be several services representing different domains that may have semantically similar type of input parameters (i.e. they are mutually *g-subsumption* matches over the *I-cluster space*). For an example the 4th cluster in the figure equally represents 4 domains: *Medical*, *Food*, *Weapon*, and *Education*. The overall *domain-recall* for both the *O-cluster space* and the *I-cluster space* is relatively low because the OWLS-TC expert classification of services into domains are primarily based on service functionality but more on the thematic category of the services. In other words, there may be services that have quite different functionality although they can pertain to the same domain thematically. For an example, *car price-lookup service* and *car rental service* generate different output (i.e. functionality) while they belong to the same domain *Economy* (i.e. services related to price).

**Query-based Accuracy Measuring:** Another alternative approach to evaluate the precision/recall of the STC algorithm is by using a set of queries. This approach resolves the inadequacy of domain based accuracy measures by strictly restricting the accuracy evaluation over the set of services that are retrieved as functionally similar to a given query (i.e. evaluating the process of service discovery itself). All such matching services are called *relevant services*. Moreover, this approach neatly evaluates the service discovery performance of the STC as a whole. We first define *Query-Precision* and *Query-Recall* as two measures for understanding the accuracy of a discovery process and hence, the goodness of a cluster space generated by a given service category learning algorithm.

**Definition 10**: *Query-Precision* with respect to a given query $Q$ is defined as the ratio of the number of relevant services that are retrieved (say $n_{rel,Q}$) when $Q$ is mapped over the cluster space generated by a service category learning algorithm $L$ vs. the total number of retrieved services ($N_{ret,Q}$) for $Q$. Numerically this means: $Pr(Q) = n_{rel,Q} / N_{ret,Q}$. ■

**Definition 11**: *Query-Recall* with respect to a given query $Q$ is defined as the ratio of the number of relevant services that are retrieved (say $n_{rel,Q}$) when $Q$ is mapped over the cluster space generated by a service category learning algorithm $L$ vs. the total number of relevant services ($N_{rel,Q}$) for $Q$. Numerically this means: $Re(Q) = n_{rel,Q} / N_{rel,Q}$. ■

To evaluate the *Query-Precision* and *Query-Recall* of STC we used the query set given in OWLS-TC v2 dataset. The query set contains 29 queries each accompanied by its corresponding expert-evaluated set of relevant services. We first calculated the average 11 point interpolated precision over recall (see definition 14) for each of the 29 queries. This accuracy measure helps us to evaluate the precision as well as the recall in an integrated manner. The underlying idea is to rank the retrieved services in order of relevancy to a given query and then to analyze each of the ranked services one by one by measuring the precision observed and the recall till that rank with respect to the given relevant set of services for that query. The precision measure is formally known as *Query-Precision@r*. We define it as follows:

**Definition 12:** *Query-Precision@r* (denoted $Pr(Q, r)$**)** Given a query $Q$ and its set $R_Q$ of relevant services ($N_{rel,Q}$), if a service discovery process over a cluster space generated by a learning algorithm $L$ retrieves an ordered set of services $R_{ret,Q}$ (cardinality, say, $N_{ret,Q}$) where the order is in decreasing sequence of relevancy of member services to $Q$, then the *Query-Precision @ r* is defined as the ratio of number of services in subset $R^r_{ret,Q} = \{s_1, s_2, ..., s_r\}$ that are members of $R_Q$ (say, $N^r_{rel,Q}$) over the total number of services (i.e. $r$) in $R^r_{ret,Q}$.■

The *Query-Precision @ r* is numerically calculated as: $Pr(Q, r) = N^r_{rel,Q} / r$.

Similarly, the corresponding *Query-Recall @ r* is calculated as: $Re(Q, r) = N^r_{rel,Q} / N_{rel,Q}$. Note here that $r$ is a positive integer such that $r = [1, N_{ret,Q}]$. $r$ is often called the *cut-off* point.

**Definition 13:** The *Interpolated Query-Precision @ r* (denoted as $I_Pr(Q,r)$) is defined at a certain recall level r = [1, 11] as the highest precision found for any recall level $r' \geq r$ $I_Pr(Q,r) = Max_{r' \geq r} Pr\ (Q,r')$. ■

**Definition 14:** The *Mean Interpolated Query-Precision@r* (denoted as $I_Pr(r)$) at a certain recall level r = [1, 11] is defined as the averaged interpolated *Query-Precision @ r* over the total number of queries ($N_Q$) that have been mapped over the cluster space by the discovery algorithm:

$$I_Pr\ (r) = \frac{\sum_{Q=1}^{N_Q} I_Pr\ (Q, r)}{N_Q} \blacksquare$$

We first observed the performance of *STC* over the *O-cluster space* with respect to each of the 29 OWLS-TC queries (O-cluster space query accuracy). The underlying service discovery process has been stratified into two phases: (i) Phase 1 where the discovery takes place only over the *O-cluster space* and (ii) Phase 2 where the discovery takes place over the *I-cluster space* after the completion of Phase 1. The accuracy evaluation involved in Phase 1 of the service discovery process is called *O-cluster space query* while that involved in Phase 2 is called *I-cluster space query*. To evaluate we calculated the average of the *Mean Interpolated Query-Precision* (average of I_Pr(Q, r) over the 11 points) for each of the queries (Fig. 12). We observed a significant increase in the overall average interpolated precision after the Phase 2 was done. This was because a considerable number of services that were identified as functionally similar to the desired output part of the queries during Phase 1 were eliminated after Phase 2 since those services could not be called either directly by the set of queries (by providing the required input consumed by those services) or indirectly by some other services that can be called themselves by the queries. This phenomenon supports the fact that service discovery should be carried as a two-phase process supported by the proposed concept of feature stratification.

TABLE I
SUMMARIZED DESCRIPTIONS OF COMPARED LEARNING ALGORITHMS

| Algorithm | Approach | Distance Measure |
|---|---|---|
| *WIC* (Word-IC algorithm) [40] | Hierarchical clustering | Common words between two services (global cluster quality). |
| *Woogle* [5] | Hierarchical clustering | Concept based distance measure *concept match* of Input/Output parameter terms. |
| *CL* (Complete Link) [24]<br>GA (Group Average) [24] | Variation of the hierarchical agglomerative hierarchical clustering | GA is an improvisation over CL. |
| *CT* (Common Term) [24] | Similar to K-means clustering | Cosine similarity – the centroid selection shared by all the clusters (cf. average of cluster members in k-means). |
| KNN (K-Nearest Neighbor) [9] | Online classification | Euclidean distance measure – computes the *k* nearest existing samples in the cluster space. |

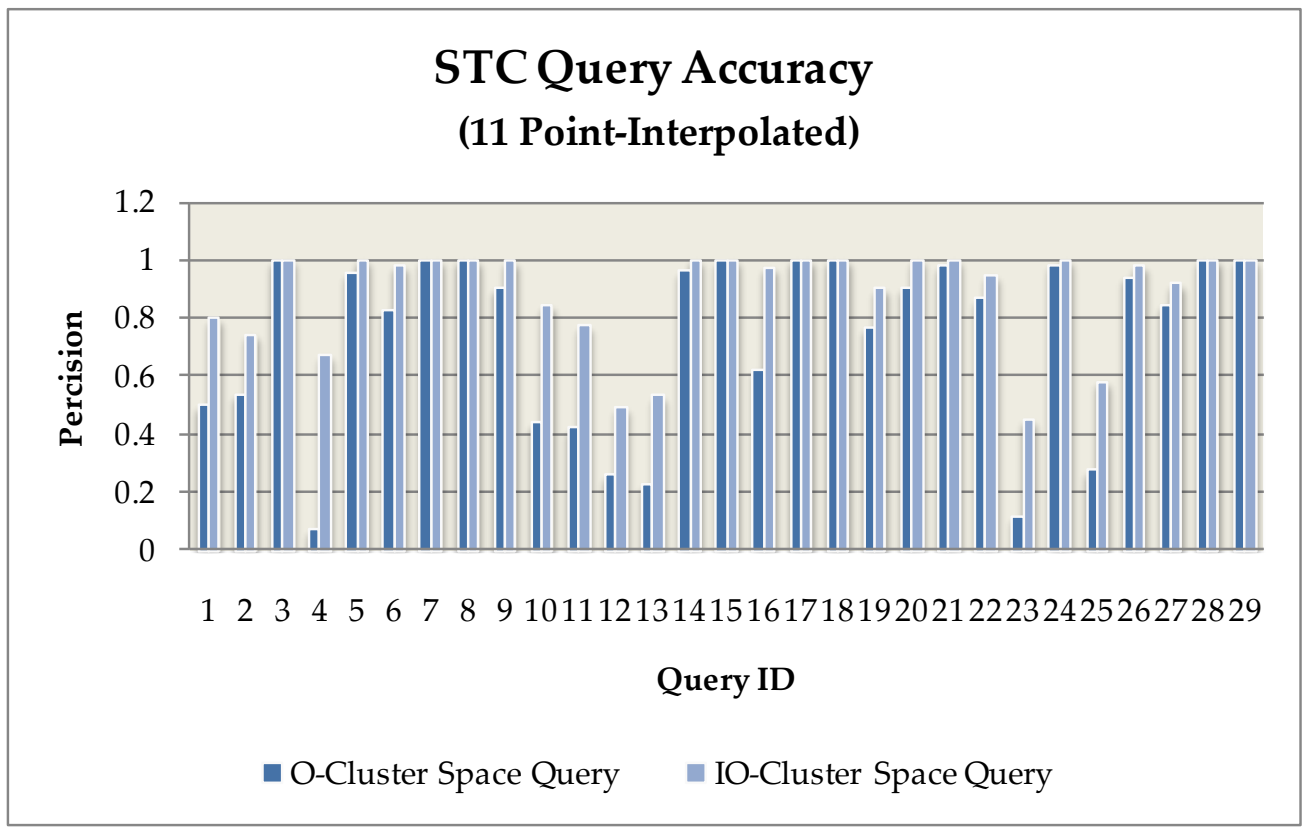

Fig. 12. Mean Interpolated Query-Precision of STC

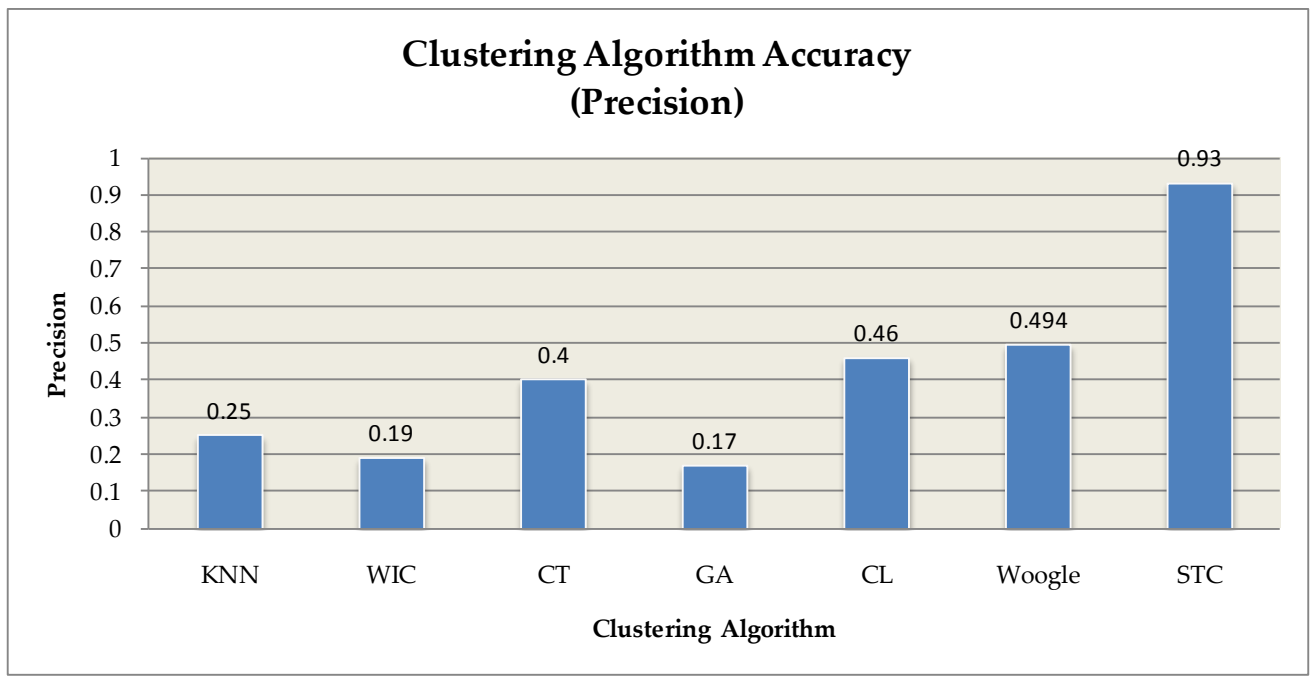

Fig. 13. Comparative Analysis of Avg. Mean Interpolated Query-Precision

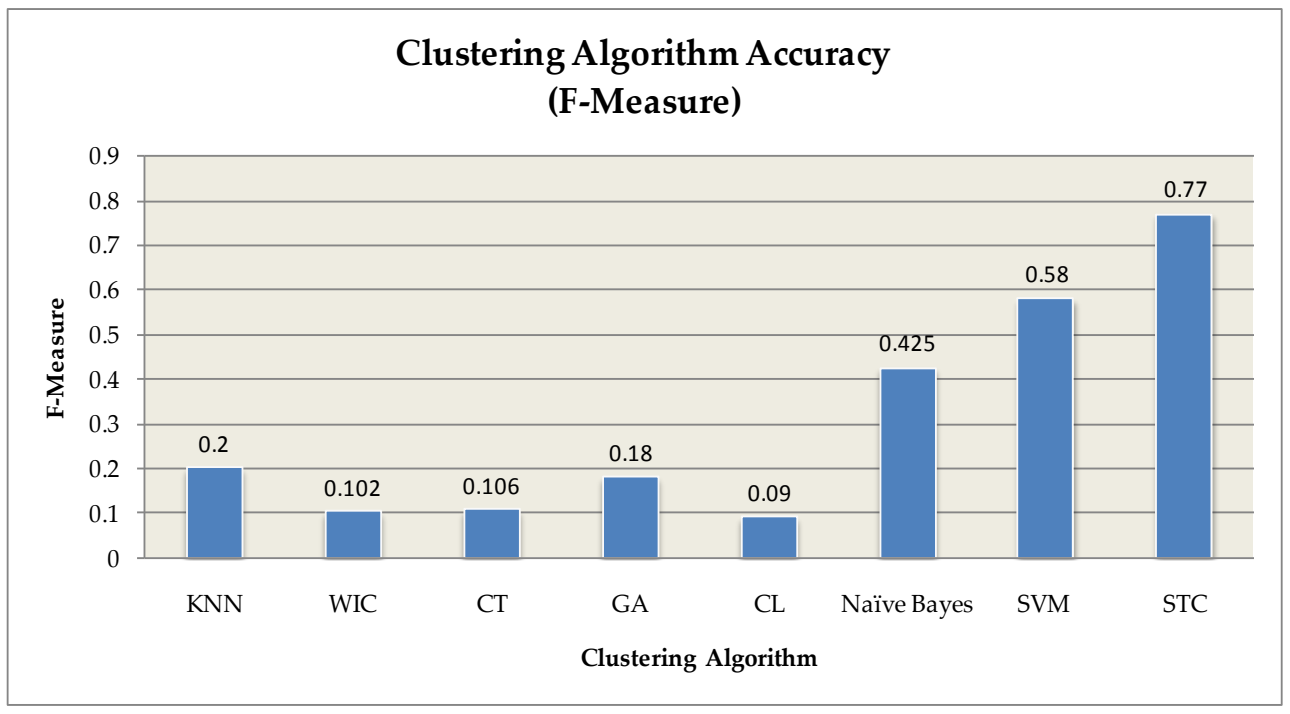

Fig. 14. Comparative Analysis of F-measure

We then compared the *Average Mean Interpolated Query-Precision* (this is *Mean Interpolated Query-Precision* when averaged over all the 29 queries) with that of six other prominent service discovery algorithms (Fig. 13). Each of them has been outlined in Table I. When we compared the F-score results of *STC* with 5 of the previous 6 algorithms (Woogle has been excluded due to lack of data) we found that it outperformed all the five algorithms again. We included the results of two supervised learning algorithms - *Naive Bayes* and *SVM (ensemble)* within this study. We observed that the F-measure score did not come out well for all the five unsupervised learning algorithms (Fig. 14). This is because their individual recall was not good enough when compared to their precision. In comparison to all these eight learning (supervised & unsupervised) algorithms we found that *STC* has a significant edge in terms of *Average Mean Interpolated Query-Precision* over all of these algorithms.

We also compared the *average 11 point interpolated precision @ r* (i.e. *I_Pr(r)* defined previously) versus the *average recall @ r* of STC-based discovery algorithm with that of 5 different service discovery techniques proposed in the OWL-S MX [33] (Fig. 15). The objective behind this study was to understand how our proposed *g-subsumption* matchmaking performed when compared to other matchmaking techniques that have used the same OWL-S TC v2 dataset as we did. The approaches M0-M4 are the different types of query match algorithms compared by OWL-S MX.

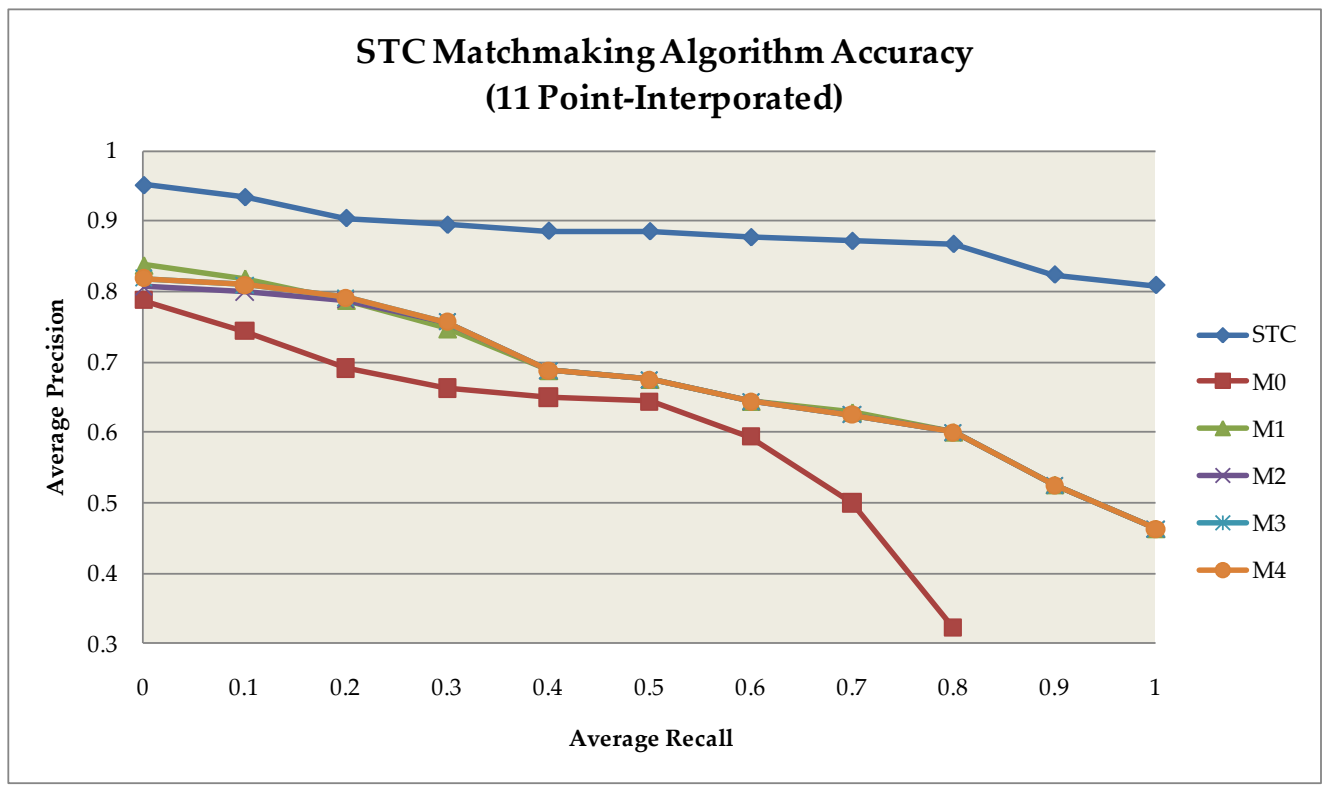

Fig. 15. Average Interpolated Query-Precision and Recall

M0 is a pure Description Logics [32] based matching algorithm that considers only the semantic definitions of the *Input/Output* parameter terms. M1 through M4 are hybrid matchmaking techniques that use both semantic definitions of parameters as well as tokens recovered from textual descriptions of services. M1 makes use of loss of information measures (LOI), M2 uses extended Jacquard similarity coefficient [41], M3 uses the cosine similarity values [42], and M4 uses the Jensen-Shannon information divergence based similarity values [43]. We found that *STC* had a significant improvement over all these matchmaking measures even when textual descriptions were not included into the service feature by *STC* (unlike M1-M4). The reason for *STC* having much better accuracy performance in comparison to M0 is that although both of them are purely based on subsumption matching of parameters, the clustering technique that uses M0 is based on ad-hoc comparison with the innate assumption that clusters are mutually disjoint. This falsely excludes services that may have a subsumption match with member services in multiple clusters. Also the case of sibling matching (e.g., *car rental* and *bus rental*) is not accounted for in M0-M4 matching. This again falsely splits services into separate disjoint clusters. Moreover, M0 being based on the Paolucci order of matching allows false inclusion of services within clusters as strong matches. This is because of higher universal preference of the *Plug-in* match over the *Subsume* match that assumes the match strength order is preserved for both the *Input* and the *Output* features. In addition, we also validate the *STC* algorithm in terms of *cluster entropy* when the service set to be clustered was confined to the relevant sets given for each query within the OWLS-TC v2 dataset and we observed average entropy of 0.19551 over 29 relevant sets (the details in Appendix).

## VI. Discussion

There are several reasons for the relatively higher accuracy of STC compared to the eight learning algorithms. STC is not based on statistical learning (unlike the others) and hence, the generalization error is not dependent on the learning parameter estimation technique but rather dependent on the semantic descriptions of the services. With classifiers such as Naïve Bayes and SVM, we need to enumerate a priori the categories in the training set that can generalize over the domain of services. In an open SOA based system the number of classes that needs to be predefined is very hard to estimate. Hence, a suitable training set is very difficult to create and that leads to unsatisfactory accuracy. On the other hand STC does not need any prior estimation of the category numbers. Furthermore SVM is still not suitable for multi-category classification.

Partitional clustering techniques such as CT have the intrinsic problem of choice of $k$ cluster value as well as the optimal initial selection of $k$ centroids. This induces sub-optimal accuracy. STC being non-partitional such problems are not applicable to it. One of the most intrinsic problems with hierarchical clustering techniques such as WIC, Woogle, CL and GA is that the estimation of the optimal level that maximizes inter-cluster distance and minimizes intra-cluster distance simultaneously is computationally expensive. KNN can be a good choice for online learning. However since it does not revise the category of existing services as new services are seen therefore the overall accuracy can be adversely affected. All the eight compared algorithms assume that the underlying service categories are disjoint. As mentioned earlier in section III, the assumption has adversely affected the individual algorithm's accuracy. In comparison STC is capable of grouping services that may belong to multiple categories.

The proposed STC algorithm has certain limitations that need to be explicated in order to completely understand its scope. First and foremost, STC works only for services that have semantic descriptions (such as OWL-S). Hence, the accuracy of the algorithm will highly depend on the correctness and completeness of the semantic service descriptions as given in the OWL-S TC test set. As OWL-S TC is not without the problems of lack of semantic richness and completeness in terms of expressivity hence, integration of efficient IR matchmaking techniques and statistical techniques for semantic definition verification into STC is currently a work-in-progress.

Secondly, the current version of STC is incapable of learning services that are semantically described by dynamically defined complex concepts (that are derived from existing ontology concepts). For an example, the output of a vehicle rental service that provides either *sedan* or *SUV* can be dynamically described using the derived concept $sedan \sqcup SUV$ where both *sedan* and *SUV* are existing ontology concepts. In such a case this complex concept will not have a corresponding *b-code* and hence, *g-relation* based comparison will not be possible. We are currently working on dynamic generation of *b-code*s for dynamically built complex concepts.

## VII. Conclusion

The paper proposes STC (Semantic Taxonomical Clustering) - an ontology encoding based clustering algorithm for automated service category learning. STC has some very significant advantages over other conventional service category algorithms since: (i) it supports multiple cluster membership, (ii) it eliminates centroid selection problem that is innate in partitional clustering algorithms, (iii) it supports online clustering, (iv) it eliminates problems of distance-based learning algorithms (hence, the problem of integrated similarity measure) by following stratified clustering approach based on non-metric *g-relation* based matchmaking, (v) it is computationally efficient with an empirically studied amortized comparison of approximately 3% of the services existing within the cluster space in an online learning framework, (vi) it has significantly highly F-measure of 0.77 (compared to the results of other approaches) as evaluated over OWL-S TC v2 data set, and (vii) its soundness and completeness (in terms of accuracy) have been proved.

## References

[1] T. Bellwood et al.. *UDDI Spec Technical Committee Draft*. 2004, Version 3.0.2. Available from http://www.uddi.org/pubs/uddi_v3.htm.

[2] I. Stoicay, R. Morrisz, D. Liben-Nowellz,D.R. Kargerz, M. F. Kaashoekz, et al, "Chord: A Scalable Peer-to-Peer Lookup Protocol for Internet Applications", *IEEE/ACM Transactions on Networking*, vol. 11, no. 1, pp. 17-32, 2003.

[3] United Nations Standard Products and Service Code. Available at http://www.unspsc.org.

[4] The North American Industry Classification System. Available at http://www.census.gov/eos/www/naics/

[5] X. Dong, A. Halevy, J. Madhavan, E. Nemes, and J. Zhang, "Similarity Search for Web Services", *Proceedings of the 13th International Conference on Very Large Data Bases*, Toronto, Canada, 2004.

[6] M. Klusch, B. Fries, and K. Sycara, "Automated Semantic Web Service Discovery with OWLS-MX", *Proceedings of the International Conference on Autonomous Agents and Multiagent Systems, 2006*.

[7] R. Lara, M. A. Corella, and P. Castells, "A Flexible Model for Web Service Discovery", *Proceedings of the 1st International Workshop on Semantic Matchmaking and Resource Retrieval* (*SMR-06*), Korea, 2006.

[8] M.A. Corella, and P. Castells, "A Heuristic Approach to Semantic Web Services Classification", *10th International Conference on Knowledge-Based & Intelligent Information & Engineering Systems* (*KES*). Universities of Brighton and Bournemouth, UK, 2006.

[9] C. Platzer, F. Rosenberg, and S. Dustdar, "Web Service Clustering using Multidimensional Angles as Proximity Measures", *ACM Transaction on Internet Technology*, vol. 9, no. 3, pp. 54-79, 2009.

[10] M. Paolucci, T. Kawamura, T. Payne, and K. Sycara, "Semantic Matching of Web Services Capabilities", *the First International Semantic Web Conference on the Semantic Web*. Sardinia, Italy, 2002.

[11] G. Wang, D. Xu, Y. Qi, and D. Hou, "A Semantic Match Algorithm for Web Services Based on Improved Semantic Distance", *4th International Conference on Next Generation Web Services Practices*, Korea, 2008.

[12] H. Yang, S. Liu, P. Fu, H. Qin, and L. Gu, "A Semantic Distance Measure for Matching Web Services", *Proceedings of the International Conference on Computational Intelligence and Software Engineering*, (*CiSE 2009*), Wuhan, China, 2009.

[13] D. Bianchini, V. Antonellis, B. Pernici, and P. Plebani, "Ontology-based Methodology for e-Service Discovery", *ACM Journal of Information Systems*, vol. 31, no. 4, pp. 361 – 380, 2006.

[14] R. Rada, H. Mili, E. Bicknell, and M. Blettner, "Development and Application of a Metric on Semantic Nets", *IEEE Transactions on Systems, Man, and Cybernetics*, vol. 19, no. 1, pp. 17–30. 1989.

[15] G. Hirst, and D. St-Onge, "Lexical chains as representations of context for the detection", *WordNet: An Electronic Lexical Database (Language, Speech, and Communication)*, The MIT Press, pp. 305–332, 1995.

[16] P. Resnik, "Using information content to evaluate semantic similarity", *Proceedings of the 14th International Joint Conference on Artificial Intelligence*, Montreal, Canada, 1995.

[17] D. Lin, "An information-theoretic definition of similarity", *Proceedings of the 15th International Conference on Machine Learning*, USA, 1998.

[18] C. Keßler, M. Raubal, and K. Janowicz, "The Effect of Context on Semantic Similarity Measurement", *On the Move to Meaningful Internet Systems: OTM 2007 Workshops*, Vilamoura, Portugal, 2007.

[19] A. Borgida, T. Walsh, and H. Hirsh, "Towards measuring similarity in description logics", *Proceedings of the 2005 International Workshop on Description Logics (DL2005)*, Edinburgh, Scotland, 2005.

[20] J. Hau, W. Lee, and J. Darlington, "A semantic similarity measure for semantic Web Services", *Proceedings of 2005 Web Service Semantics Workshop*. Chiba, Japan, 2005.

[21] H. Wang, Y. Shi, X. Zhou, Q. Zhou, S. Shao, and A. Bouguettaya, "Web Service Classification using Support Vector Machine", *22nd IEEE International Conference on Tools with Artificial Intelligence (ICTAI)*, Arras, France, 2010.

[22] M. Bruno, G. Canfora, M.D. Penta, and R. Scognamiglio, "An Approach to Support Web Service Classification and Annotation", *Proceedings of the IEEE International Conference on e-Technology, e-Commerce and e-Service* (*EEE'05*), Washington DC, USA, 2005.

[23] N. Oldham, C. Thomas, A. Sheth, and K. Verma, "METEOR-S Web Service Annotation Framework with Machine Learning Classification", *International Conference on Semantic Web Services and Web Process Composition*, San Diego, USA, 2004.

[24] A. Heß, and N. Kushmerick, "Learning to Attach Semantic Metadata to Web Services", *Proceedings of the 2nd International Semantic Web Conference (ISWC 2003)*, Sanibel Island, USA, 2003.

[25] A. Sajjanhar, J. Hou, and Y., Zhang, "Algorithm for Web Services Matching", *Proceedings of the 6th Asia-Pacific Web Conference (APWeb '04)*, Hangzhou, China, 2004.

[26] J. Ma, J. Cao, Y. Zhang, "A Probabilistic Semantic Approach for Discovering Web Services", *Proceedings of the 2007 IEEE World Wide Web Conference*, Banff, Canada, 2007.

[27] J. Ma, J. Cao, Y. Zhang, "Efficiently Finding Web Services Using a Clustering Semantic Approach", *Proceedings of the 2008 International Workshop on Context Enabled Source and Service Selection, Integration and Adaptation (CESSS '08)*, Beijing, China, 2008.

[28] L-J. Zhanga, S. Chengb, Y-M. Cheea, A. Allamc, Q. Zhoua, "Pattern Recognition based Adaptive Categorization Technique and Solution for Services Selection", *Proceedings of the IEEE Asia-Pacific Services Computing Conference (APSCC '07)*, Tsukuba, Japan, 2007.

[29] Q. Liang., P. Li, P.C.K. Hung, and X. Wu, "Clustering Web Services for Automatic Categorization", *Proceedings of the 2009 IEEE International Conference on Services Computing (SCC '09)*, Bangalore, India, 2009.

[30] C. Godfrey and A.W. Siddons, "*Modern Geometry (page 20)*". Cambridge University Press, London, UK, 1908.

[31] M. Burstein, J. Hobbs, O. Lassila, D. Mcdermott, S. Mcilraith, S. Narayanan, M. Paolucci, B. Parsia, T. Payne, E. Sirin, N. Srinivasan, and K. Sycara, "OWL-S: Semantic Markup for Web Services", November 2004, W3C recommendation.

[32] F. Baader, I. Horrocks, and U. Sattler, "*Handbook of Knowledge Representation*" (*Chapter 3 – Description Logics*). Elsevier, 2007.

[33] M. Klusch, B. Fries, and K. Sycara, "OWL-S-MX: A hybrid Semantic Web Service matchmaker for OWL-S services", *Web Semantics: Science, Services and Agents on the World Wide Web*, 7(2):121-133, 2009.

[34] K. Sycara, S. Widoff, M. Klusch, and J. Lu, "LARKS: Dynamic Matchmaking Among Heterogeneous Software Agents in Cyberspace", *Autonomous Agents and Multi-Agent Systems*, 5(2):173–203, 2002.

[35] U. Bellur, and R. Kulkarni, "Improved Matchmaking Algorithm for Semantic Web Services Based on Bipartite Graph Matching", *IEEE International Conference on Web Services* (ICWS-07), USA, 2007.

[36] S.B. Mokhtar, D. Preuveneers, N. Georgantas, V. Issarny, and Y. Berbers, "EASY: Efficient semAntic Service discoverY in pervasive computing environments with QoS and context support", *Journal of Systems and Software*, vol. 81, no. 5, pp. 785 – 808, 2008.

[37] E. Sirin, B. Parsia, B.C. Grau, A. Kalyanpur, and Y. Katz, "Pellet: A Practical OWL-DL Reasoner", *Journal of Web Semantics: Science (Services and Agents on the World Wide Web*, vol. 5, no. 2, 2007.

[38] I. Horrocks, "The FaCT System", *Automated Reasoning with Analytic Tableaux and Related Methods: International Conference Tableaux'98*, LNAI, Springer-Verlag, vol. 1397, pp. 307-312, 1998.

[39] S. Dasgupta, S. Bhat, and Y. Lee, "SGPS: A Semantic Scheme for Web Service Similarity". *Proceedings of IEEE World Wide Web Conference*, Madrid, Spain, 2009.

[40] O. Zamir, O. Etzioni, O. Madani, and R.M. Karp, "Fast and Intuitive Clustering of Web Documents". *Proceedings of the 3rd International Conference on Knowledge Discovery and Data Mining*, USA, 1997.

[41] A. Strehl, and J. Ghosh, "Value-based Customer Grouping from Large Retail Data-sets", *Proceedings of SPIE Conference on Data Mining and Knowledge Discovery*, Orlando, USA, 2000.

[42] C.D. Manning, P. Raghavan, and H. Schutze, *Introduction to Information Retrieval*, Cambridge University Press, USA, 2008.

[43] J. Lin, "Divergence Measures based on the Shannon Entropy", *IEEE Transactions on Information Theory*, vol. 37, no. 1, pp. 145–151, 1991.